\documentclass[%
preprint, superscriptaddress,
amsmath,amssymb,
aps,
]{revtex4-2}
\usepackage{xcolor}
\usepackage[colorlinks=true, linkcolor = red, citecolor = blue]{hyperref}
\usepackage{graphicx}
\usepackage{amsmath}
\usepackage{dcolumn}
\usepackage{bm}
\usepackage{etoolbox}

\makeatletter
\patchcmd{\frontmatter@abstract@produce}
  {\penalty-200\relax}
  {\penalty10000\relax}
  {}
  {}
\makeatother

\makeatletter
\let\@listcomma@comma\@listcomma@comma@UK
\makeatother

\begin{document}

\title{Black holes from a Higgs-like field in the radiation era}%
\author{Ethan Milligan}
\email{e.milligan@qmul.ac.uk}
\affiliation{Astronomy Unit, Queen Mary University of London,
Mile End Road, London, E1 4NS, UK}

\author{Luis E. Padilla}
\affiliation{Department of Physics, Rikkyo University, Tokyo, Japan}

\author{David J. Mulryne}
\affiliation{Astronomy Unit, Queen Mary University of London,
Mile End Road, London, E1 4NS, UK}

\date{\today}

\begin{abstract}
 
Light spectator fields during inflation can acquire superhorizon fluctuations that cross a potential barrier between positive and negative regions of their potential.
Motivated by the Standard Model Higgs instability, 
in this work we study the subsequent evolution 
of patches where this occurs in the radiation era after inflation ends for a Higgs-like spectator field.
We utilise fully nonlinear, spherically symmetric numerical relativity. Across the black hole forming configurations in our investigation we find a robust two-stage evolution. First, the central negative potential region reverses its expansion, becomes kinetic dominated, and forms a primordial black hole that hides the runaway core. The positive potential barrier that survives outside this first horizon then determines one of two late-time branches. In the subcritical branch the original apparent horizon grows smoothly and engulfs the remaining scalar structure. In the supercritical branch, however, the potential energy of the barrier dominates the local evolution. The result is a transient wormhole throat, a bifurcating trapping horizon, and an inflating child-universe branch. In both branches the parent radiation dominated universe is ultimately left with an ordinary primordial black hole whose subsequent growth is governed by radiation accretion. 
\end{abstract}

\maketitle

\clearpage
\section{Introduction}
Scalar fields with metastable potentials arise frequently in high energy physics.
Examples include extensions of the Standard Model, effective descriptions of high energy physics, and the string theory landscape \cite{Kachru2003,Balasubramanian2005,Douglas2003}. 
If such a field is light during inflation, stochastic fluctuations can displace it over a potential barrier between stable and unstable parts of the potential in some regions of the universe. The subsequent fate of these patches is a genuinely nonlinear gravitational problem: the patch may explore a region with negative potential energy, while the surrounding cosmology continues to evolve normally.

The Standard Model Higgs provides an especially well known example. Favoured values for the masses of the Higgs and top quark suggest that the Standard Model Higgs potential becomes metastable at large field values \cite{Degrassi2012,Buttazzo2013}. 
This means that if the Higgs field is light during inflation, over-the-barrier fluctuations may occur in this setting \cite{StarobinskyYokoyama1994,Enqvist2014}. 
We must then ask: \emph{What is the fate of the resulting patches?} The answer to this question impacts directly on whether Higgs metastability poses a catastrophic threat to our universe, or instead gives rise to localised strong gravity phenomena such as primordial black holes (PBHs).

Initially, it was generally assumed that regions of the universe that fluctuate over the barrier would act like anti-de Sitter bubbles and expand outwards at the speed of light \cite{PhysRevD.15.2929, PhysRevD.21.3305}. After inflation ends, even one bubble within our past light cone would then be sufficient to destroy our universe \cite{Espinosa:2015qea, Espinosa_2008, DeLuca:2022cus, Fairbairn_2014, Kobakhidze:2013tn, Hook:2014uia, Kohri_2016}. More recently there has been some debate about the fate of such patches. It was realised that since the total energy density in these regions of the universe is initially likely to be positive, they will at first evolve like a patch of FRW spacetime. Later, however, they will eventually start to collapse, as the field evolves to more negative regions of the potential and in turn a singularity forms at the centre of the region with a horizon around it. Taking this into account, work based on a thin wall approximation for spherically symmetric regions suggested that all of the unstable region of the potential is finally hidden behind a horizon and  the result is a black hole that does not threaten the universe \cite{DeLuca:2022cus}. Later studies  using numerical relativity claimed, however, that this is not the case \citep{Strumia:2022kez}. They argued that the thin wall approximation is not appropriate for Higgs patches generated by stochastic excursions over the barrier and going beyond this approximation found in their simulations that part of the unstable region is still exposed to the outside universe. This region will grow, like in the original AdS bubble picture,  restoring the original ``Higgstory" \cite{Espinosa:2015qea} (see also \citep{East:2016anr} for earlier numerical relativity simulations with a similar conclusion ).

The Higgs instability 
with stochastic fluctuations is one example of a broader class of gravitational phenomena.  Scalar field regions separated from their surroundings by a closed domain wall like transition region also arise in other settings. For example, another important example occurs in cosmology with axion fields, where all vacua are stable. Literature focusing on collapse of a spherically symmetric domain wall in this setting, in the post-inflationary era, shows that the nonlinear gravitational dynamics that develops is richer than a simple choice between expansion and collapse \citep{Deng:2016vzb,Deng:2017uwc,ning2026numericalsimulationsprimordialblack}, as we discuss below. Motivated by the Higgs instability literature and the ubiquitousness of meta-stable potentials, in this work we study the post-inflationary evolution of a field with a metastable Higgs-like potential,  and question the fate of over-the-barrier patches.
  
We utilise fully nonlinear, spherically symmetric numerical relativity simulations. 
In contrast to previous Higgs instability work (but in common with work on axion fields \citep{Deng:2016vzb,Deng:2017uwc,ning2026numericalsimulationsprimordialblack}), our simulations are based in the radiation era after inflation ends. This is motivated by the fact that for a wide range of inflationary energy scales the initial collapse is likely to occur only after reheating. Moreover, even if the collapse did commence during inflation, any expanding region would only threaten the universe once it ends (since before then inflation ensures that even regions growing at the speed of light become causally disconnected from the bulk of the universe). Our simulations therefore include gravity, a scalar field and radiation.  

\emph{To summarise our results, we find, importantly,  that the final state is always a standard black hole embedded in a radiation dominated FRW background, with any unstable over-the-barrier region entirely hidden behind horizons. }

As in the axion case mentioned above, the precise form of  evolution which leads to this final state is varied and complex. In particular, it depends sensitively on the  ratio of the initial physical size of the barrier region to its self-gravity length (defined in Section~\ref{section:LT}). The ratio measures whether the barrier region of the spacetime is ``subcritical" or ``supercritical". In the supercritical case, after an initial black hole forms due the collapse of the region with negative potential energy, the barrier region drives the spacetime towards a throat configuration. The derivative of the areal radius vanishes there and a second trapped surface eventually appears outside the original horizon. In the geometric terminology used in the PBH literature, our initial data are Type-I-like because the initial areal radius is monotonic. Nevertheless, the supercritical branch dynamically develops a throat and a bifurcating trapping
horizon, placing the late-time solution in the Type-B class. The Type-B structure is therefore generated by the nonlinear evolution of the scalar-field barrier rather than inherited from an initially Type-II geometry \cite{Kopp2011,Uehara_2025,shimada2024primordialblackholeformation}. Geometrically, the interpretation is that an  inflating inner region driven by the hilltop of the potential surrounds an inner black hole and is connected by a wormhole throat to the parent radiation dominated FRW universe. The bifurcation of the marginal surface then marks the pinching off of the throat and the baby universe containing a black hole  becomes causally disconnected from the surrounding FRW region, which sees only the second outer horizon. Hence the negative potential region remains hidden, rather than spreading into exterior cosmology and external observers see only an ordinary black hole.

By contrast, in the subcritical case the evolution exhibits a Type-A trapping horizon structure. The initial collapse produces a black hole enclosing the negative potential region, and the subsequent barrier evolution is not strong enough to trigger the secondary geometric transition associated with the supercritical Type-B branch. Even in this case, however, the black hole grows to enclose the entire over-the-barrier region. Moreover, in both cases the subsequent growth of the apparent horizon is consistent with accretion from the surrounding radiation background. The two cases are highly analogous to the super and subcritical branches discussed in the axion literature, with the important difference that, in our case, the stage driven by the hilltop region begins only after a negative potential region black hole has already formed due to the negative region of the potential. 

We now detail the numerical set-up that leads to these results, the assumptions we make and the simulations which lead us to these conclusions.

\section{Numerical Simulation}
\subsection{Einstein-Klein-Gordon system of equations}
We perform our numerical evolution in a spherically symmetric setting, using a metric of the form
\begin{equation}
    ds^2 = -A(r,t)^2dt^2 + B(r,t)^2dr^2 + R(r,t)^2d\Omega^2\,.
\end{equation}
Here $R(r,t)$ is the areal radius, which we assume to vanish at the origin of spherical coordinates, $B(r,t)$ is the radial metric component and $d\Omega^2\equiv d\theta^2 + \sin^2\theta d\varphi^2$. We exploit the gauge freedom in the choice of lapse function by choosing the geodesic slicing. This corresponds to the choice $A(r,t)=1$ and implies that the coordinate time $t$ is the proper time of observers at fixed spatial coordinates. We consider Einstein gravity minimally coupled to a canonical scalar field and a perfect fluid, defined by the  action
\begin{equation}
    S = \int d^4x\sqrt{-g}\bigg[\frac{^{(4)}R}{16\pi} - \frac{1}{2}g^{\mu\nu}\partial_\mu h\partial_\nu h - V(h) + \mathcal{L}_{\text{fluid}}\bigg]\,,
\end{equation}
where $^{(4)}R$ is the 4D Ricci scalar, $g_{\mu\nu}$ is the metric tensor and $h$ is a scalar field with a potential $V(h)$ and $\mathcal{L}_{\text{fluid}}$ is the Lagrangian density of a perfect fluid. The energy momentum tensors of the scalar field and fluid are
\begin{subequations}
\begin{equation}
    T^{(h)}_{\mu\nu} = \partial_\mu h\,\partial_\nu h - g_{\mu\nu}\left(\frac{1}{2}g^{\alpha\beta}\partial_\alpha h \partial_{\beta} h + V(h)\right)\,,
\end{equation}
\begin{equation}
T^{(\text{fluid})}_{\mu\nu}=(\rho + p )u_{\mu}u_{\nu} + p g_{\mu\nu}\,,
\end{equation}
\end{subequations}
where $\rho$, $p$ and $u^{\mu}$ are the fluid energy density, pressure and 4-velocity, respectively. The fluid equation of state is taken to be $p=w\rho$ and throughout we consider a radiation fluid and hence $w=1/3$. The fluid's 4-velocity can be written in the form
\begin{equation}
    u^{\mu}(r,t) = \bigg(\frac{1}{A\sqrt{1 - v^2}},\frac{v}{B\sqrt{1 - v^2}},0,0\bigg)\,,
\end{equation}
where $v(r,t)$ is the fluid's 3-velocity relative to the comoving coordinate $r$. 

For numerical convenience, we introduce dimensionless variables by rescaling with the Hubble parameter at the end of inflation $H_i$,
\begin{equation}
\label{eq:rescaled}
\begin{aligned}
r &\to \frac{r} {H_i}, \qquad
t \to \frac{t} {H_i}, \qquad
\rho \to M_{ \rm Pl}^2 H_i^2 \rho,
\qquad h \to M_{ \rm Pl}h, \qquad
V \to M_{ \rm Pl}^2 H_i^2 V,
\end{aligned}
\end{equation}
where $M_{ \rm Pl} = 1/\sqrt{G}$, so that all variables become dimensionless. From this point onward, unless explicitly stated otherwise, the symbols $r$, $t$, $\rho$, $h$ and $V$ denote the corresponding dimensionless quantities defined by Eq.~\eqref{eq:rescaled}. Following \citep{Cho_1997,bloomfield2015formalismprimordialblackhole,Deng:2016vzb} we introduce
\begin{equation}
    U \equiv \dot R,
    \qquad
    \Gamma \equiv \frac{R'}{B},
    \qquad
    K \equiv \frac{\dot B}{B}+2\frac{\dot R}{R},
\end{equation}
where $\dot{}\equiv\partial/\partial t$ and
$'\equiv\partial/\partial r$. Thus $\dot R=U$ and
$\partial_rR=B\Gamma$, relations that will also be used below as geometric and numerical diagnostics. The Einstein equations together with the conservation of the stress energy tensors give \citep{Deng:2016vzb,ning2026numericalsimulationsprimordialblack}
\begin{widetext}
\begin{subequations}\label{eq:system}
\renewcommand{\theequation}{\theparentequation.\arabic{equation}}
\begin{align}
\dot{K} &=
-\left( K - \frac{2U}{R} \right)^2
- 2\left( \frac{U}{R} \right)^2
- 4\pi \left( T_{00} + T^{1}{}_{1} + 2T^{2}{}_{2} \right)\,,
\label{eq:system1}
\\[0.6em]
\dot{U} &=
-\frac{1-\Gamma^2+U^2}{2R}
- 4\pi R T^{1}{}_{1}\,,
\label{eq:system2}
\\[0.6em]
\dot{\Gamma} &=
-\frac{4\pi R T^{0}{}_{1}}{B}\,,
\label{eq:system3}
\\[0.6em]
\dot{\rho} &=
\frac{(1+w)\rho}{1-wv^2}
\left[
v^2\left( K-\frac{2U}{R} \right)
- K
- \frac{2v\Gamma}{R}
- \frac{v'}{B}
\right]
-\frac{1-w}{1-wv^2}\frac{\rho'v}{B}\,,
\label{eq:system4}
\\[0.6em]
\dot{v} &=
\frac{(1-v^2)v}{1-wv^2}
\left[
-\left( K-\frac{2U}{R} \right)
+ wK
+ \frac{2wv\Gamma}{R}
-\frac{(1-v^2)w}{(1+w)v}\frac{\rho'}{\rho B}
\right]
-\frac{1-w}{1-wv^2}\frac{v'v}{B}\,,
\label{eq:system5}
\\[0.6em]
\ddot{h} &=
-K\dot{h}
+\frac{1}{BR^2}\left( \frac{R^2}{B}h' \right)'
-\partial_h V\,,
\label{eq:system6}
\\[0.6em]
\dot{B} &= B\left( K-\frac{2U}{R} \right)\,,
\label{eq:system7}
\\[0.6em]
\dot{R} &= U\,,
\label{eq:system8}
\end{align}
\end{subequations}
\end{widetext}
where the stress-energy components are 
\begin{subequations}\label{eq:stress_tensor}
\begin{align}
T_{00} &=
\frac{1+w v^2}{1-v^2}\rho
+\frac{1}{2}\dot{h}^{\,2}
+\frac{1}{2B^2}h'^2
+V(h)\,,
\label{eq:stress_tensor_a}
\\[0.6em]
T^{1}{}_{1} &=
\frac{w+v^2}{1-v^2}\rho
+\frac{1}{2}\dot{h}^{\,2}
+\frac{1}{2B^2}h'^2
-V(h)\,,
\label{eq:stress_tensor_b}
\\[0.6em]
T^{2}{}_{2} &=
w\rho
+\frac{1}{2}\dot{h}^{\,2}
-\frac{1}{2B^2}h'^2
-V(h)\,,
\label{eq:stress_tensor_c}
\\[0.6em]
T^{0}{}_1 &=
\dot{h}\,h'
-\frac{1+w}{1-v^2}\rho v B\,.
\label{eq:stress_tensor_d}
\end{align}
\end{subequations}

\subsection{Initial Conditions}
We model the metastable spectator field using the near-barrier approximation to the Higgs potential employed in previous studies of Higgs vacuum instability \citep{Espinosa_2008,Strumia:2022kez,Espinosa:2015qea,East:2016anr}. In physical variables we write
\begin{equation}
    V_{ \rm phys}(h_{\rm phys})
    \simeq
    -b_{ \rm phys}
    \ln\!\left(
        \frac{h_{\rm phys}^2}
             {h_{\max,{\rm phys}}^2\sqrt{e}}
    \right)
    \frac{h_{\rm phys}^4}{4},
    \label{higgs_potential_physical}
\end{equation}
where $h_{\max, {\rm phys}}$ is the field value at the maximum of the potential and
\begin{equation}
    b_{\rm  phys}=\frac{0.16}{(4\pi)^2},
    \qquad
    V_{\max}=\frac{b_{\rm phys}}{8}h_{\max, {\rm phys}}^4.
\end{equation}
The dimensionless potential used in the numerical evolution is obtained by applying the rescalings of Eq.~\eqref{eq:rescaled}. In this work we use the Higgs form of the potential but do not restrict its overall field scale to the value corresponding to the physical Standard Model Higgs.

The physical picture motivating our scalar initial conditions is that the field behaves as a light spectator during inflation and is subsequently evolved classically from the beginning of radiation domination. We denote the Hubble scale at this time by $H_i$. The corresponding physical background energy density is
\begin{equation}
    \rho_i=\frac{3H_i^2}{8\pi G}
    =\frac{3M_{ \rm Pl}^2H_i^2}{8\pi},
\end{equation}
where $M_{ \rm Pl}=G^{-1/2}$ in our conventions. The stochastic origin of the initial profile requires the scalar to be light during inflation,
\begin{equation}
    m^2_{\text{eff}}\equiv |V^{\prime\prime}(h_{ \rm max})| \ll H_i^2,
\end{equation}
and that the scalar energy density is subdominant compared with an inflationary background $\rho_h\ll\rho_i$. Both conditions are satisfied by the representative configurations considered below.

We take the initial scalar configuration to be a smooth Gaussian profile centred at $r=0$, with the central field value beyond the maximum of the potential. At large radius the field approaches a non-zero background value $h_{ \rm eq}$. This is motivated by the equilibrium distribution of a light scalar during stochastic inflation as on the scale of the observable Hubble patch, $h_{ \rm eq}$ represents the background value about which rarer fluctuations on shorter scales occur. We assume $h_{ \rm eq}<h_{\max}$ so that a localised over-the-barrier fluctuation is embedded in a surrounding metastable region. We have also verified that setting $h_{ \rm eq}=0$ does not change the qualitative collapse behaviour described below.

Approximating the potential away from the hilltop by a quartic, the stationary stochastic distribution gives \citep{StarobinskyYokoyama1994,Enqvist:2013kaa,Markkanen:2018pdo}
\begin{equation}
    \langle h^2\rangle
    \simeq
    0.132\,H_i^2\lambda^{-1/2},
\end{equation}
and we therefore take the asymptotic field value to be of order
\begin{equation}
    h_{ \rm eq}
    \sim
    \sqrt{\langle h^2\rangle}
    \simeq
    0.363\,H_i\lambda^{-1/4},
    \label{eq:heq}
\end{equation}
noting that the precise value depends on the detailed form of the potential.

\begin{figure}[t]
    \centering
    \includegraphics[width=0.70\linewidth]{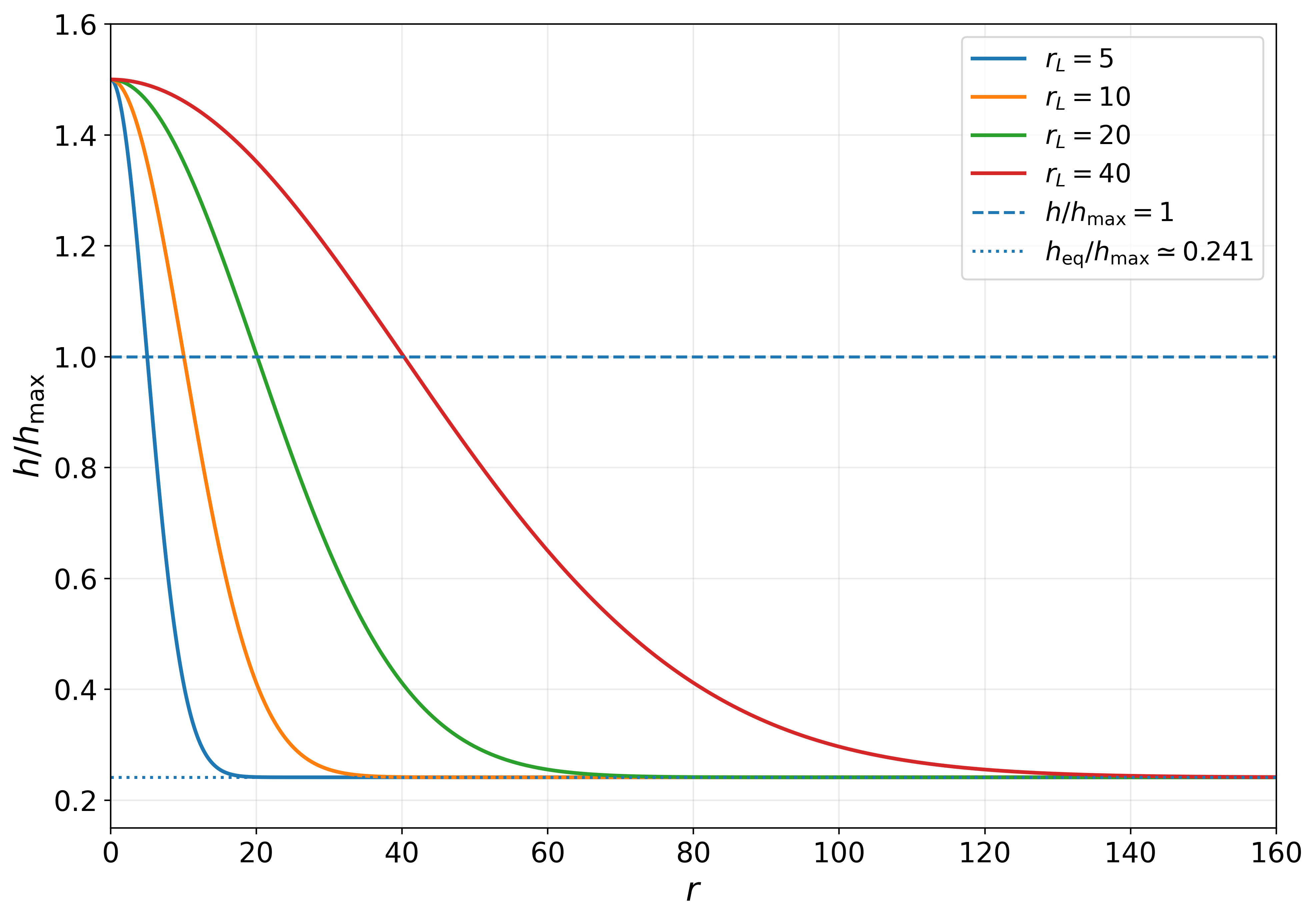}
    \caption{\footnotesize{
    Representative initial scalar field profiles described by
    Eq.~\eqref{eq:initial_higgs_profile} for different superhorizon
    lengthscales $r_L$. The central field value is fixed at
    $h_i=1.5h_{\max}$, while the field asymptotes to $h_{ \rm eq}$ at  large radius. The dashed horizontal line denotes the maximum of the potential, $h=h_{\max}$ and the dotted horizontal line denotes the asymptotic background value $h=h_{ \rm eq}$. The representative simulations discussed in the main text use $r_L=20$. Recall we are using the rescaled $r$ defined in Eq.~(\ref{eq:rescaled}).}
    }
    \label{fig:initial_higgs_profiles}
\end{figure}

The initial spatial profile is
\begin{equation}
    h(r)
    =
    (h_i-h_{ \rm eq})
    e^{-r^2/(2r_L^2)}
    +h_{ \rm eq},
    \label{eq:initial_higgs_profile}
\end{equation}
so that $h(0)=h_i$ and $h(r\rightarrow\infty)=h_{ \rm eq}$. Here $r_L$ is the dimensionless width defined using the rescaled radial coordinate of Eq.~\eqref{eq:rescaled}.  All configurations considered in the main analysis are initially superhorizon, $r_L\gg1$. Representative initial profiles for different values of $r_L$ are shown in Fig.~\ref{fig:initial_higgs_profiles}.

For the remaining initial data we embed the scalar profile in a spatially flat radiation dominated FRW geometry. In the dimensionless variables of Eq.~\eqref{eq:rescaled},
\begin{equation}
    B(r,t_0)=1,
    \qquad
    R(r,t_0)=r,
    \qquad
    \Gamma(r,t_0)=1,
    \qquad
    v(r,t_0)=0,
    \qquad
    \dot h(r,t_0)=0.
\end{equation}
Following Ref.~\citep{Deng:2016vzb}, we compensate the initial scalar energy with a perturbation in the radiation density,
\begin{equation}
    \rho(r,t_0)
    =
    \frac{3}{8\pi}
    -\frac{1}{2}\left(\partial_rh\right)^2
    -V(h),
\end{equation}
so that the total initial energy density is homogeneous, $T_{00}=3/(8\pi)$. The Einstein constraint equations are then satisfied by
\begin{equation}
    K(r,t_0)=3,
    \qquad
    U(r,t_0)=r,
\end{equation}
corresponding to the initial FRW expansion. These compensated isocurvature initial conditions are a convenient way of keeping the initial geometry exactly spatially flat FRW. They require a compensating correlation between the scalar and radiation perturbations, which we do not expect to arise generically from the inflationary dynamics, and should therefore be regarded
as a convenient choice of initial data rather than as a physical prediction. They should not be interpreted as requiring a physical correlation between the inflationary Higgs and radiation perturbations. To test the robustness of our results, we have therefore also performed evolutions beginning with an initially homogeneous radiation density, solving the Einstein constraints for the corresponding inhomogeneous geometry. We recover the same qualitative subcritical and supercritical formation branches, showing that our results do not rely on the compensated radiation prescription.

\subsection{Subcritical and Supercritical Diagnostics}
\label{sec:branchdefinitions}

Before specifying the representative numerical values of $H_i$, $h_{\max}$ and $r_L$, we introduce the quantities that determine the late-time evolution. This is important because the parameter choices below are made specifically to span both of the nonlinear branches found in the simulations. The simulations reveal two qualitatively different late-time evolutions in the region of the positive potential barrier that survives outside the first black hole, which we refer to as subcritical and supercritical. Similar branches occur in the gravitational dynamics of spherical domain walls and vacuum bubbles \citep{Deng:2016vzb,Deng:2017uwc,Blau,Aurilia:1989sb}. In those systems, the competition between the physical size of the wall and the gravitational lengthscale associated with its tension determines whether the wall remains gravitationally weak or instead produces a wormhole and an inflating child-universe.

Our scalar configurations are broad, dynamical profiles rather than thin walls, but the same physics motivates a measure of the self-gravity of the surviving barrier. For a thin wall with surface energy density $\sigma$, the corresponding gravitational length is
\begin{equation}
    \ell_\sigma
    =
    \frac{1}{2\pi G\sigma},
\end{equation}
motivated by the classic gravitating-wall solutions \cite{Vilenkin:1981zs,Vilenkin:1983wt,Ipser:1983db}. For the broad scalar profile we replace the thin-wall tension by an effective excess scalar energy \cite{Ghassemi2007,Guerrero:2002ki},
\begin{equation}
    \sigma_{\rm eff}(t_0)
    \equiv\int_{\mathcal B_w}d\ell\, \left[\rho_h(t_0,\ell)-\rho_{h, \rm bg}(t_0)\right],
    \qquad
    d\ell=B(t_0,r)dr,
    \label{eq:sigma_eff}
\end{equation}
where $\mathcal B_w$ denotes the radial region containing the positive potential barrier. $\rho_{\rm h}$ is the scalar field energy density and $\rho_{h, \rm bg}$ is its asymptotic background value in the surrounding metastable region. Thus, this computes the positive energy density excess of the barrier. We then define
\begin{equation}
    \ell_{\sigma, {\rm eff}}
    \equiv
    \frac{1}{2\pi G\sigma_{ \rm eff}}.
    \label{eq:ellsigmaeff}
\end{equation}
In the dimensionless variables used in the numerical evolution $G$ has been scaled out, reducing this expression to the form used below in the simulations.

To quantify the spatial scale of the barrier we define its radius $R_w$ by the outward crossing of the maximum of the potential,
\begin{equation}
    h(t_0,r_w)=h_{\max},
    \qquad
    R_w=R(t_0,r_w).
    \label{eq:Rwdef}
\end{equation}
For the compensated initial conditions the initial hypersurface is spatially flat, so $R(t_0,r)=r$. Substituting the Gaussian profile of Eq.~\eqref{eq:initial_higgs_profile} therefore gives
\begin{equation}
    R_w = r_L \sqrt{ 2\ln\!\left(\frac{h_i-h_{ \rm eq}} {h_{\max}-h_{ \rm eq}} \right)}.
    \label{eq:Rw_rL}
\end{equation}
Thus, for fixed $h_i/h_{\max}$ and $h_{ \rm eq}/h_{\max}$, the ratio $R_w/r_L$ is fixed: changing $r_L$ simply rescales the physical radius of the barrier.

The corresponding initial barrier compactness is
\begin{equation}
    \left.
    \frac{R_w}{\ell_{\sigma, \rm eff}}
    \right|_{t_0} = 2\pi G\sigma_{ \rm eff}(t_0)R_w.
    \label{eq:Cw}
\end{equation}
This quantity compares the physical radius of the barrier with the gravitational length associated with the scalar energy stored across it, and therefore provides an initial measure of the barrier self-gravity. For the Gaussian spatial profile used here, the dependence of this quantity on the initial physical parameters can be seen without evaluating the detailed profile integral. Let $\Delta_w$ denote the physical radial width of the positive potential barrier and let $\rho_{h,w}$ denote its characteristic excess scalar energy density. Equation~\eqref{eq:sigma_eff} gives approximately
\begin{equation}
    \sigma_{ \rm eff}
    \sim
    \rho_{h,w}\Delta_w.
    \label{eq:sigma_scaling}
\end{equation}
For the broad profiles relevant to the supercritical transition, the characteristic scalar energy scale in the barrier is set by the height of the potential,
\begin{equation}
    \rho_{h,w}
    \sim
    V_{\max}
    =
    \frac{b_{\rm phys}}{8}h_{\max, {\rm phys}}^4.
\end{equation}
Moreover, when $h_i/h_{\max}$ and $h_{ \rm eq}/h_{\max}$ are fixed, both the barrier width and its radius scale with the physical Gaussian width,
\begin{equation}
    \Delta_w\propto r_{L, {\rm phys}},
    \qquad
    R_w\propto r_{L, {\rm phys}}.
\end{equation}
Combining these relations with Eq.~\eqref{eq:Cw} gives
\begin{equation}
    \left.
    \frac{R_w}{\ell_{\sigma, {\rm eff}}}
    \right|_{t_0}
    \sim
    G b_{ \rm phys}h_{\max, {\rm phys} }^4r_{L, {\rm phys}}^2,
\end{equation}
where the numerical coefficient associated with the precise Gaussian shape and the operational definition of $\mathcal B_w$ has been suppressed. Using $G=M_{ \rm Pl}^{-2}$ and $r_L=H_i r_{L, \rm phys}$, this becomes
\begin{equation}\label{eq:Rw_scaling}
    \left.
    \frac{R_w}{\ell_{\sigma, {\rm eff}}}\right|_{t_0}
    \simeq
    b_{ \rm phys}\left(\frac{h_{\max, {\rm phys}}}{H_i}\right)^4\left(\frac{H_i}{M_{ \rm Pl}}\right)^2 r_L^2.
\end{equation}
Equation~\eqref{eq:Rw_scaling} is intended as the approximate scaling of the exact initial diagnostic in Eq.~\eqref{eq:Cw}. All quoted numerical values of $R_w/\ell_{\sigma, {\rm eff}}$ below are obtained from the full profile. The scaling makes clear that the initial self-gravity depends not only on the local energy scale of the potential but also on the spatial extent of the fluctuation. In particular, at fixed $h_{\max}/H_i$, decreasing $H_i/M_{ \rm Pl}$ suppresses the initial barrier compactness, whereas increasing $r_L$ enhances it quadratically.

A second quantity becomes important dynamically. We use the local,
time-dependent ratio
\begin{equation}
    \frac{V[h(t,r)]}{\rho_{\rm fluid}(t,r)}
\end{equation}
to determine whether the positive part of the scalar potential becomes locally important relative to the radiation background. The sign of this ratio is crucial. Since $\rho_{\rm fluid}>0$, a positive ratio identifies a region with $V>0$, whereas a negative ratio corresponds to the unstable $V<0$ region. In particular,
\begin{equation}
    \frac{V}{\rho_{\rm fluid}}\gtrsim1
\end{equation}
indicates that the positive potential dominates over the local radiation density. When the scalar kinetic and gradient contributions are also subdominant, the stress-energy becomes sufficiently vacuum energy-like to support local accelerated expansion, approaching $p\simeq-\rho$, as the potential becomes increasingly dominant. This allows for a period of local accelerated expansion. By contrast, if $V/\rho_{\rm fluid}<-1$ the potential is negative and its magnitude exceeds the local radiation density. This does not correspond to a locally inflating region.

The two diagnostics therefore play different roles. In every supercritical evolution studied here, the initial barrier compactness satisfies
\begin{equation}
    \left.\frac{R_w}{\ell_{\sigma,\rm eff}}\right|_{t_0}
    \gtrsim 1,
\end{equation}
while $V/\rho_{\rm fluid}$ is initially much smaller than unity. The $V/\rho_{\rm fluid}\gtrsim1$ condition develops only during the subsequent evolution. As the surviving positive potential barrier evolves in the exterior close to the growing black hole apparent horizon, the local radiation density decreases through cosmological redshifting and accretion onto the black hole. The positive barrier therefore becomes increasingly important relative to the local radiation and evolves to satisfy $V/\rho_{\rm fluid} > 1$ before the secondary geometric transition develops.

By contrast, the subcritical configurations have weaker initial barrier
self-gravity,
\begin{equation}
    \left.\frac{R_w}{\ell_{\sigma,\rm eff}}\right|_{t_0}
    \lesssim1,
\end{equation}
and the positive potential barrier does not develop an extended region with
$V/\rho_{\rm fluid}\gtrsim1$ while it remains dynamically relevant in the
exterior.

We therefore regard $R_w/\ell_{\sigma, {\rm eff}}$ as an initial predictor of the late-time branch within the family of configurations studied here, whereas $V/\rho_{ \rm fluid}\gtrsim1$ characterises the dynamical state of the barrier when the supercritical transition actually occurs. The order-unity values should be understood as empirical diagnostics of the configurations studied here rather than as universal critical thresholds.

\subsection{Choice of Representative Initial Parameters}
\label{sec:parameterchoice}

Having defined the quantities that organise the late-time evolution, we now choose representative values for the free parameters in the Gaussian initial profile. Following previous studies of Higgs vacuum instability, we characterise the initial fluctuation by its field amplitude relative to the potential barrier and its characteristic lengthscale relative to the cosmological scale \cite{Espinosa:2015qea,Strumia:2022kez}.

Defining
\begin{equation}
    \epsilon
    \equiv
    \frac{H_i}{M_{ \rm Pl}},
\end{equation}
the representative configurations used in the main text satisfy
\begin{equation}
    \frac{h_{\max, \rm phys}}{H_i}=4,
    \qquad
    \frac{h_i}{h_{\max}}=1.5,
    \qquad
    r_L=H_i r_{L, \rm phys}=20.
    \label{eq:scaledparams_physical}
\end{equation}
Equivalently, in the dimensionless field variables used in the evolution,
\begin{equation}
    h_{\max}=4\epsilon,
    \qquad
    h_i=1.5h_{\max}=6\epsilon,
    \qquad
    r_L=20.
    \label{eq:scaledparams}
\end{equation}
The relation $h_{\max}/H_i=4$ is a choice defining the representative Higgs-like profile studied here. It should not be interpreted as a physical relation fixing the position of the Standard Model Higgs maximum in terms of the inflationary Hubble scale. For a fixed Standard Model potential, $h_{\max}$ is determined independently, and varying $H_i$ changes both the
stochastic production of over-the-barrier patches and their gravitational importance relative to the radiation background. At fixed $h_{\max}/H_i$ and $r_L$, Eq.~\eqref{eq:Rw_scaling} shows that varying $H_i/M_{ \rm Pl}$ changes the initial self-gravity of the barrier. We take
\begin{equation}\label{eq:lower}
    \frac{H_i}{M_{ \rm Pl}} = 10^{-5}
\end{equation}
for the representative subcritical evolution and
\begin{equation}
    \frac{H_i}{M_{ \rm Pl}} = 10^{-1}
\end{equation}
for the representative supercritical evolution. These choices allow both nonlinear mechanisms to be resolved using the same dimensionless profile shape.

These values should not be interpreted as observationally motivated
inflationary scales for the physical Standard Model Higgs.  In particular, the larger value is used to make the strongly self-gravitating branch numerically accessible at the moderate superhorizon scale $r_L=20$. The simulations should instead be viewed as a controlled exploration of the nonlinear dynamics of a Higgs-like metastable potential.

In terms of the lightness of the spectator, at the maximum of the potential in Eq.~\eqref{higgs_potential_physical},
\begin{equation}
    |V''(h_{\max})|
    =
    2b_{ \rm phys}h_{\max}^2,
\end{equation}
so that
\begin{equation}
    \frac{|V''(h_{\max})|}{H_i^2}
    =
    2b_{ \rm phys}
    \left(\frac{h_{\max}}{H_i}\right)^2.
\end{equation}
For the representative choice $h_{\max}/H_i=4$ this gives
\begin{equation}
    \frac{|V''(h_{\max})|}{H_i^2}
    =32b_{ \rm phys}
    \simeq3.2\times10^{-2}<1,
\end{equation}
so the field remains light for both representative configurations. Varying $H_i/M_{ \rm Pl}$ is therefore being used to vary the gravitational importance of the barrier, not to move between light- and heavy-spectator regimes. The scaling in Eq.~\eqref{eq:Rw_scaling} also shows how the location of the two branches changes as the overall scale is lowered. The supercritical branch is therefore not eliminated as the inflationary scale is lowered. Instead, at fixed $h_{\max}/H_i$ it is displaced toward increasingly extended superhorizon configurations, with the required lengthscale growing as $H_i/M_{ \rm Pl}$ decreases. The numerical challenge at low inflationary scales is consequently a
reflection of this physical scaling rather than evidence for the disappearance of the supercritical branch. The representative values used here should therefore be viewed as a numerically accessible way of demonstrating both nonlinear mechanisms for a Higgs-like potential.

The stochastic inflationary picture provides a natural interpretation of these field amplitudes. For an approximately constant Hubble rate, each e-fold of inflation produces a stochastic kick of typical size, 
\begin{eqnarray}
    \delta h \simeq \frac{H_i}{2\pi}.
\end{eqnarray}
Therefore, over a finite interval of $N$ e-folds, the root-mean-square fluctuation of a light field is
\begin{eqnarray}\label{eq:rms}
    \sigma_N\simeq \frac{H_i}{2\pi}\sqrt{N}.
\end{eqnarray}
Here $\sigma_N$ denotes the stochastic field standard deviation accumulated over the chosen range of inflationary scales. It should not be confused with the wall tension introduced later in the discussion of the barrier self-gravity. For the particular physical mapping adopted above, Eqs.~\eqref{eq:heq}, \eqref{eq:rms} and \eqref{eq:scaledparams} give an initial excursion of
approximately $5\sigma_N$ for $N=50$. This estimate provides a stochastic motivation for considering a rare over-the-barrier peak. For a different physical mapping of the same numerical solution, the corresponding stochastic rarity will change.

The usefulness of this stochastic estimate is that it allows us to quantify how rare an over-the-barrier patch is. The central value of our profile should be thought of as a rare realisation drawn from the stochastic distribution generated during the observable number of e-folds of inflation about the background equilibrium value.
We can ask how many standard deviations away from the equilibrium value the Higgs must fluctuate in order to exceed the maximum of the potential. The number is important for two reasons. First, it measures the rarity of the patches whose subsequent evolution we simulate. Second, rare high peaks of a statistically homogeneous and isotropic random field are expected to be well approximated by the spherically averaged profile around the peak \citep{Bardeen:1985tr, Yoo:2018esr, Germani:2025sphericity}. If the required excursion were only of order one standard deviation, for example, over-the-barrier regions would both be common and not appear as isolated spherically symmetric objects embedded in an otherwise standard radiation dominated universe.

\subsection{Expansions and Horizons}\label{EandH}
For a spatial 2-sphere $\mathcal{S}$ in any spacetime, there are two future-directed null directions normal to $\mathcal{S}$ corresponding to the outgoing ($+$) and ingoing ($-$) radial null congruences. The associated expansion $\Theta$ measures the rate of change of the area of $\mathcal{S}$ along a given null direction and thus describes the expansion or contraction of a bundle of null rays. If $\Theta<0$, the area of $\mathcal{S}$ decreases in that direction and the null rays converge; if $\Theta>0$, the area increases and the null rays diverge. We denote the expansions of the outgoing and ingoing null rays by $\Theta^+$ and $\Theta^-$, respectively. In flat spacetime, one has $\Theta^-<0$ and $\Theta^+>0$, corresponding to the convergence of ingoing null rays and the divergence of outgoing null rays. Such surfaces are referred to as normal surfaces. By contrast, if both expansions are negative  $\Theta^\pm<0$ the surface is said to be trapped, as occurs inside a black hole. If both expansions are positive, $\Theta^\pm>0$, the surface is anti-trapped, as occurs for sufficiently large spheres outside the cosmological apparent horizon in an expanding FRW spacetime.

In our coordinate system, the null vectors are
\begin{equation}
    k^\pm_\mu = \frac{1}{\sqrt{2}}(-A,\pm B,0,0)\,,
\end{equation}
satisfying $k^+ \cdot k^-=-1$. The expansions are defined by
\begin{equation}
    \Theta^\pm\equiv h^{\mu\nu}\nabla_\mu k^\pm_\nu\,,
\end{equation}
where $h^{\mu\nu}$ is the induced metric on $\mathcal{S}$. This reduces to,
\begin{equation}
    \Theta^\pm=\frac{\sqrt{2}}{R}(U\pm\Gamma)\,.
\end{equation}
Specifically, a black hole marginally trapped surface satisfies
$\Theta^+=0$ with $\Theta^-<0$. Here $M$ denotes the Misner-Sharp quasi-local mass, defined explicitly in Sec.~II~F below. Since
$\Theta^+\Theta^-=2(U^2-\Gamma^2)/R^2$, the condition $U^2=\Gamma^2$, equivalently $2M=R$, identifies a marginal surface. The black hole apparent horizon branch additionally satisfies $\Theta^+=0$ with $\Theta^-<0$, whereas the cosmological/anti-trapped branch satisfies $\Theta^-=0$ with $\Theta^+>0$. A bifurcating trapping horizon occurs when $\Theta^+=\Theta^-=0$ simultaneously.

In a flat FRW universe,
\begin{equation}
    \Theta^+\propto H + \frac{1}{R} \quad\text{and}\quad \Theta^-\propto H - \frac{1}{R},
\end{equation}
which should be satisfied in regions in the false vacuum at the outer boundary.
\subsection{Misner-Sharp Mass}
We characterise the gravitational energy contained within a sphere of areal radius $R$ using the Misner-Sharp mass \citep{Misner,Hayward_1996}. This defines the quasi-local mass of an object and is defined in spherical symmetry as,
\begin{equation}
    M = \frac{R}{2}(1 - g^{\mu\nu}\partial_\mu R\partial_\nu R).
\end{equation}
In our coordinates, this becomes
\begin{equation}\label{eq:misnersharpmass}
    M = \frac{R}{2}(1 + U^2 - \Gamma^2)\,.
\end{equation}
From the Einstein equations we obtain the radial derivative of the mass
\begin{equation}\label{eq:ham}
    M^\prime=4\pi R^2(R^\prime T_{00} - \dot{R}T_{01})\,.
\end{equation}
The above equation is a combination of the Hamiltonian and momentum constraints and we use this to check the numerical accuracy of the simulations. At each time slice we calculate $M$ from Eq.~\eqref{eq:misnersharpmass}, determine its spatial derivative using the same finite difference scheme employed in the evolution and compare this with Eq.~\eqref{eq:ham}. This is discussed in \ref{sec:appendix}.
\subsection{Numerical Implementation}
We evolve Eqs.~\eqref{eq:system1} - \eqref{eq:system8} using the method of lines. Radial derivatives are discretised using fourth-order finite difference stencils on a uniform grid of $N$ points and time integration is performed using a fourth-order Runge-Kutta integration. At the inner boundary, we add ghost cells to utilise the parity of the variables and enforce boundary conditions. The functions $A,B,\Gamma,K,h$ are even functions of $r$ and $U,R,v$ are odd functions of $r$. Ensuring the computational domain is sufficiently large, we do not specify any outer boundary conditions but use backwards finite derivatives. For radii sufficiently below the outer boundary at $r_{\text{max}}$, the spacetime approaches an FRW background containing the false vacuum. We choose $r_{\max}$ sufficiently large that no signal from the outer numerical boundary reaches the region of interest during the simulated time interval. Furthermore, we find that numerical stability is improved by adopting a grid-centred computational domain, thereby avoiding the placement of a grid point at $r=0$ \citep{Baumgarte:2010ndz}. 

Following apparent horizon formation, the region interior to the outermost marginally trapped surface is excised from the computational domain. The horizon position is identified using the null expansions as described in section \ref{EandH}. The radial grid is therefore dynamical and is reconstructed with its inner boundary at the apparent horizon. All evolution variables are interpolated onto the excised domain. Spatial derivatives near the inner boundary are evaluated using forward finite difference stencils.

To suppress high frequency numerical modes, we apply p=3 Kreiss–Oliger dissipation \citep{Kreiss_1972}. In this scheme, for all evolution variables $ f \in \{U,R,v,\rho,K,\Gamma,h,B\}$, the evolution equations are modified as follows
\begin{equation}
\begin{split}
\partial_t f_m \rightarrow \partial_t f_m
+ \frac{\sigma}{64\Delta r}\Bigl(
f_{m+3} - 6f_{m+2} + 15f_{m+1} - 20f_m
+ 15f_{m-1} - 6f_{m-2} + f_{m-3}
\Bigr)\,,
\end{split}
\end{equation}
where $m\pm n$ labels the grid point, $n$ the total offset from m and $\sigma$ is an adjustable dissipation parameter of order $\mathcal{O}(10^{-2})$. At the inner boundary, the stencil is completed with reflected ghost cells, with either even or odd parity imposed as appropriate for the variable. For simulations with a dynamically excised inner boundary the dissipation operator is modified by reducing its local stencil radius to the largest value compatible with the available grid points. 
\section{Universal first collapse: formation of the initial primordial black hole}\label{sec:firstcollapse}

In this section  we describe the geometric feature that is common to all Higgs-like initial conditions we investigate, the inner horizon, before turning in Sec. IV to the non-universal evolution of the surviving positive potential barrier. Across all Higgs-like initial conditions that form a black hole, the early evolution follows the same sequence. The over-the-barrier core first rolls into $V<0$ region. Next, the  negative energy of the potential becomes large enough to overcome the local radiation background and this causes the local expansion to become contraction.
This contraction then rapidly blueshifts the scalar kinetic energy, producing a  trapped region and the first apparent horizon. To illustrate this stage we use one representative subcritical run from Eq.~\eqref{eq:lower}
\begin{equation}\label{eq:representative_subc_ics}
    \frac{H_i}{M_{ \rm Pl}} = 1\times10^{-5},
    \qquad
    h_{\max} = 4H_i,
    \qquad
    h_i = 1.5h_{\text{max}},
    \qquad
    H_i r_L = 20 .
\end{equation}
\subsection{Energy density dynamics during the first collapse phase}\label{sec:eddynamics} 

\begin{figure}[h!]
\centering
\includegraphics[width=0.70\linewidth]{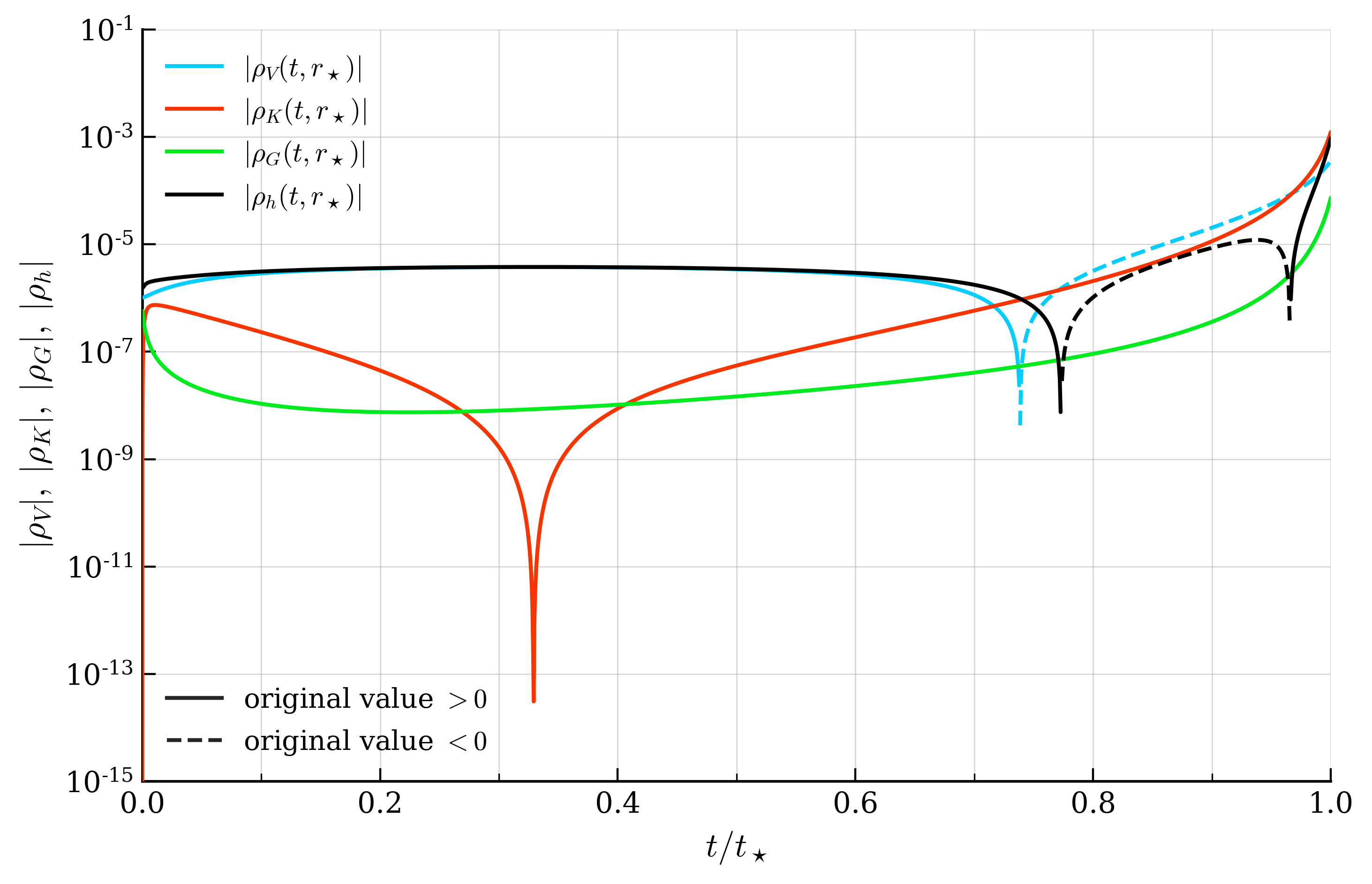}
\caption{\footnotesize{Evolution of the local Higgs field energy decomposition at $r_\star$. The vertical axis shows absolute values, while solid (dashed) segments indicate that the original quantity is positive (negative). These results were obtained from an initial superhorizon Gaussian fluctuation.}}
\label{fig:centralrhos}
\end{figure}

\begin{figure}[h!]
\centering
\includegraphics[width=0.80\linewidth]{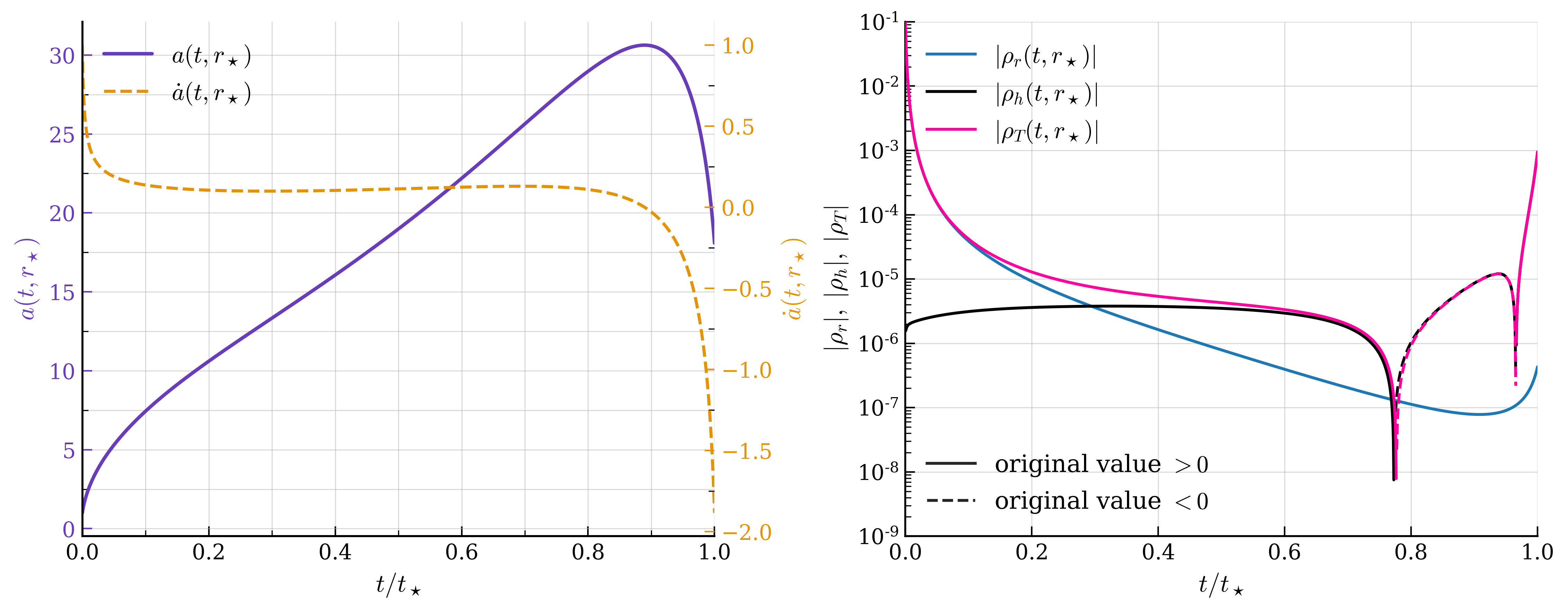}
\caption{\footnotesize{Left: evolution of the local scale factor $a(t,r_\star)$ and its time derivative at $r_\star$. The turnaround occurs when $a(t,r_\star)$ reaches its maximum and $\dot a(t,r_\star)=0$. Right: comparison between the radiation density $\rho_r$, the Higgs contribution $\rho_h$ and the total density $\rho_T=\rho_r+\rho_h$ at the same radius.}}
\label{fig:2fig_scale_total}
\end{figure}

We decompose the scalar energy density as
\begin{equation}
    \rho_h=\rho_K+\rho_G+\rho_V,
    \qquad
    \rho_K=\frac12\dot h^2,
    \quad
    \rho_G=\frac{h'^2}{2B^2},
    \quad
    \rho_V=V(h).
\end{equation}
We define $t_\star$ as the earliest time at which a marginally trapped surface satisfying $\Theta^+=0$ and $\Theta^-<0$ appears, and $r_\star$ as its coordinate radius on that slice. Figure~\ref{fig:centralrhos} shows the evolution of these contributions at $r_\star$. At early times, $t/t_\star\ll1$, the patch is locally expanding and the gradient contribution remains subdominant. Since the initial kinetic energy vanishes, $\rho_K$ begins to grow as the field rolls but this growth is suppressed by the Hubble friction term $-K\dot h$ in Eq.~\eqref{eq:system6}, leaving the potential contribution dominant. As the field rolls deeper into the unstable region, $V(h)$ becomes negative and grows in magnitude. The Higgs contribution $\rho_h$ therefore becomes negative and eventually offsets the positive radiation density. For comparison with the scalar contribution, we define the radiation contribution to the local energy density measured by observers normal to the constant-$t$ slices as
\begin{equation}
    \rho_r
    \equiv
    T_{00}^{({\rm fluid})}
    =
    \frac{1+w v^2}{1-v^2}\rho ,
\end{equation}
where $\rho$ is the radiation energy density in the fluid rest frame. The total local energy density is therefore
\begin{equation}
    \rho_T
    \equiv T_{00}
    =
    \rho_r+\rho_h .
\end{equation} 
As shown in Fig.~\ref{fig:2fig_scale_total}, this coincides with the total local density $\rho_T=\rho_r+\rho_h$ approaching zero and the local scale factor $a(t,r_\star)$ reaching its maximum, with $\dot a(t,r_\star)=0$. The patch consequently turns around from expansion to contraction. Once contraction begins, the term $-K\dot h$ acts as anti-friction, rapidly amplifying $\dot h$. The kinetic contribution then grows much faster than the potential and gradient terms, so that $\rho_h$ becomes positive again and the collapse approaches apparent horizon formation in a kinetic dominated regime. The gradient contribution remains subdominant throughout.

This establishes the energy-matter dynamical sequence preceding horizon formation. We now verify geometrically that the ensuing kinetic dominated contraction produces the first trapped region. In Sec.~IV we turn to the surviving positive potential barrier, whose properties determine the subsequent late-time evolution.

\subsection{Formation of the initial negative potential region primordial black hole}

Once the local turnaround has occurred and the Higgs patch enters the kinetic dominated contraction phase, the collapse produces a trapped region inside the unstable side of the potential. The first apparent horizon is identified by the condition $\Theta^+=0$ with $\Theta^-<0$, or equivalently by $2M=R$. In all of the Higgs-like evolutions that we studied, this marginally trapped surface forms before any late-time behaviour associated with the barrier region develops. The initial outcome of the collapse is therefore always a primordial black hole whose interior contains the crunching negative potential region.

\begin{figure}[h!]
    \centering
    \includegraphics[width=0.8\linewidth]{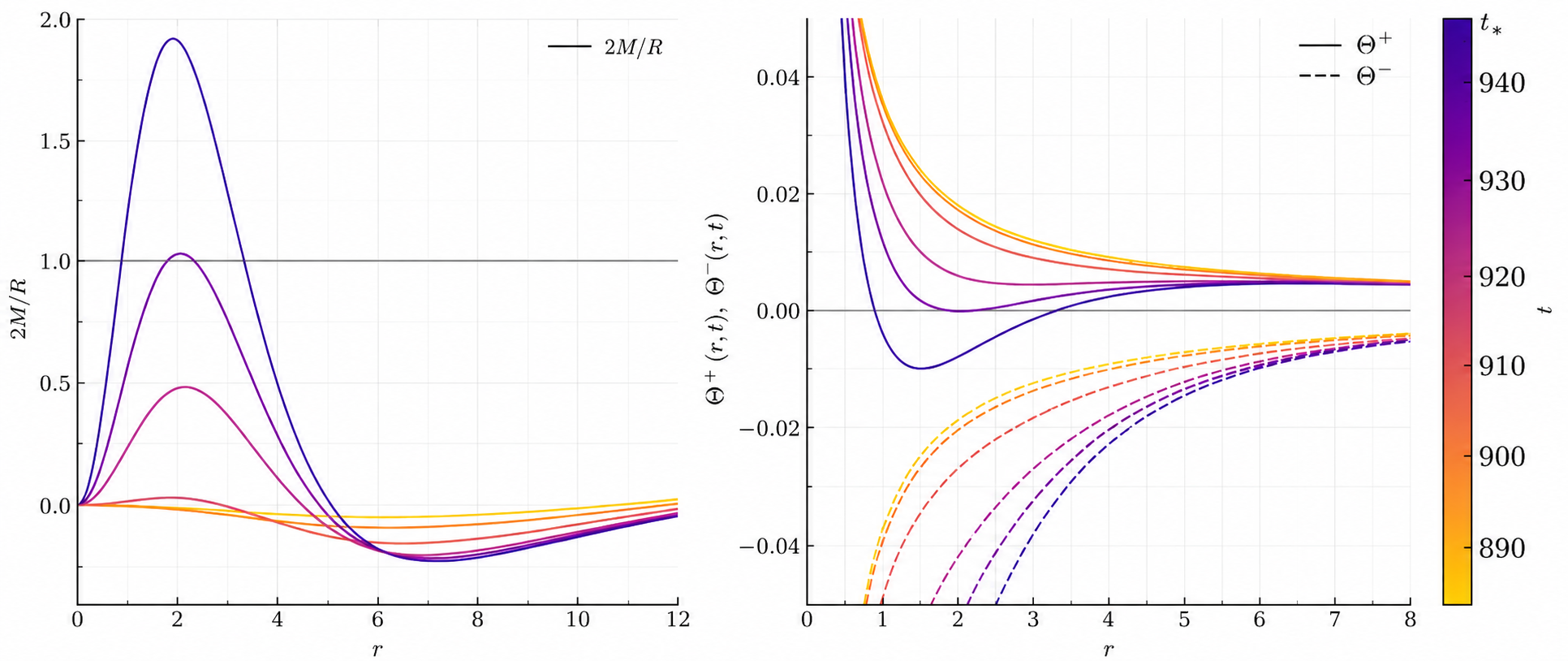}
    \caption{\footnotesize{Geometric diagnostics of the first collapse on a sequence of time slices approaching and passing through $t_\star$. \textbf{Left}: compactness $2M/R$ as a function of radius. The horizontal line marks the threshold $2M/R=1$ for apparent horizon formation. \textbf{Right}: Outgoing and ingoing null expansions, $\Theta^+$ and $\Theta^-$.}}
    \label{fig:firsthorizon}
\end{figure}

\begin{figure}[h!]
    \centering
    \includegraphics[width=0.8\linewidth]{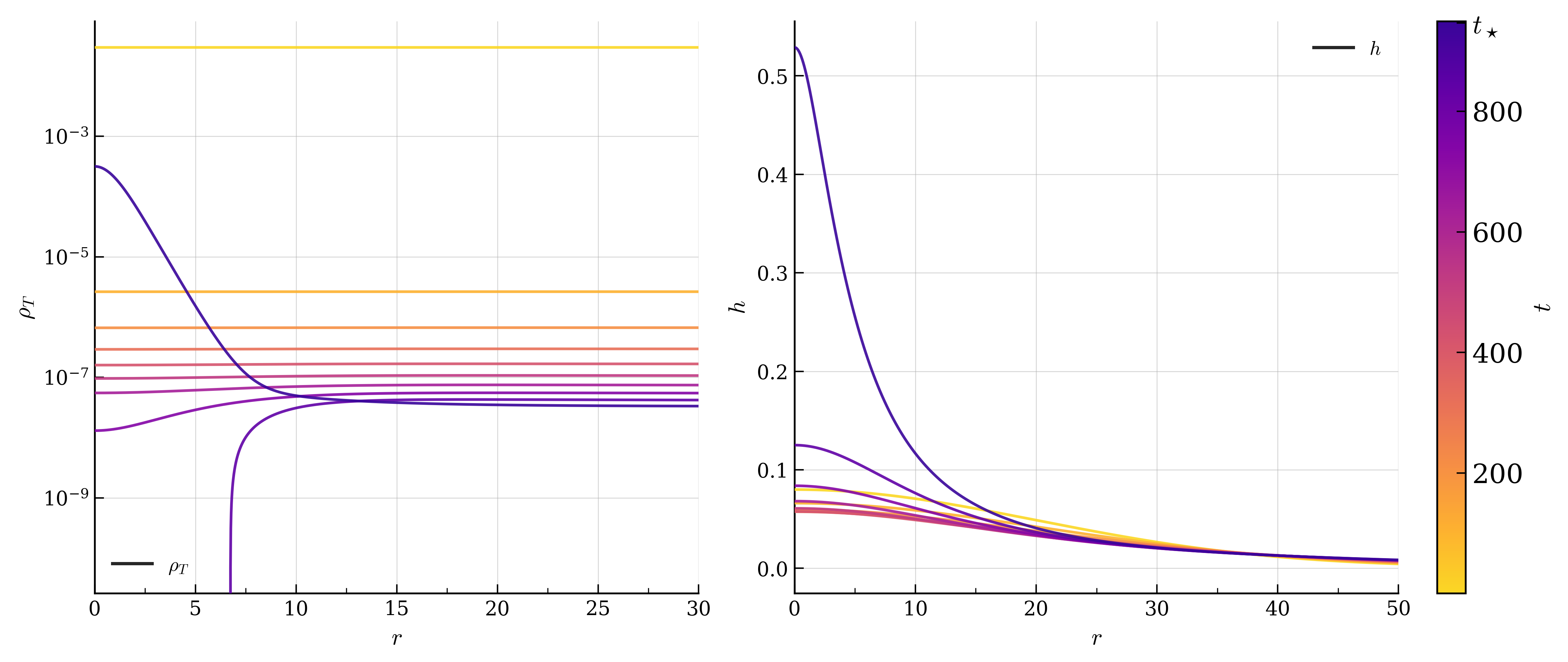}
    \caption{\footnotesize{Radial profiles on a sequence of time slices approaching the first apparent horizon formation time $t_\star$. \textbf{Left}: Total energy density $\rho_T(r)$. \textbf{Right}: Higgs profile $h(r)$.}}
    \label{fig:rhotandphi}
\end{figure}

The corresponding geometric diagnostics are shown in Fig.~\ref{fig:firsthorizon}. In the left panel we plot the compactness $2M/R$ on a sequence of time slices. Initially, $2M/R<1$ everywhere so no trapped region is present. As the central overdensity grows the compactness develops a localised peak near the origin. The apparent horizon forms at the time $t_\star$ when this peak first reaches the threshold $2M/R=1$. At later times the peak exceeds unity over a finite radial interval, indicating the presence of a trapped region surrounding the crunching negative potential region. The right panel shows the corresponding null expansions. Before horizon formation one has $\Theta^+>0$ and $\Theta^-<0$, as expected for a normal region. At $t=t_\star$, the outgoing expansion first vanishes while the ingoing expansion remains negative, identifying a marginally trapped surface. For later slices, $\Theta^+$ becomes negative inside the trapped region, while $\Theta^-$ stays negative throughout. This indicates that the first collapse produces a Type-A apparent horizon. In particular, only one zero of $\Theta^+$ is present in this stage so the initial outcome is a single apparent horizon enclosing the negative energy interior.

\begin{figure}[h!]
    \centering
    \includegraphics[width=0.8\linewidth]{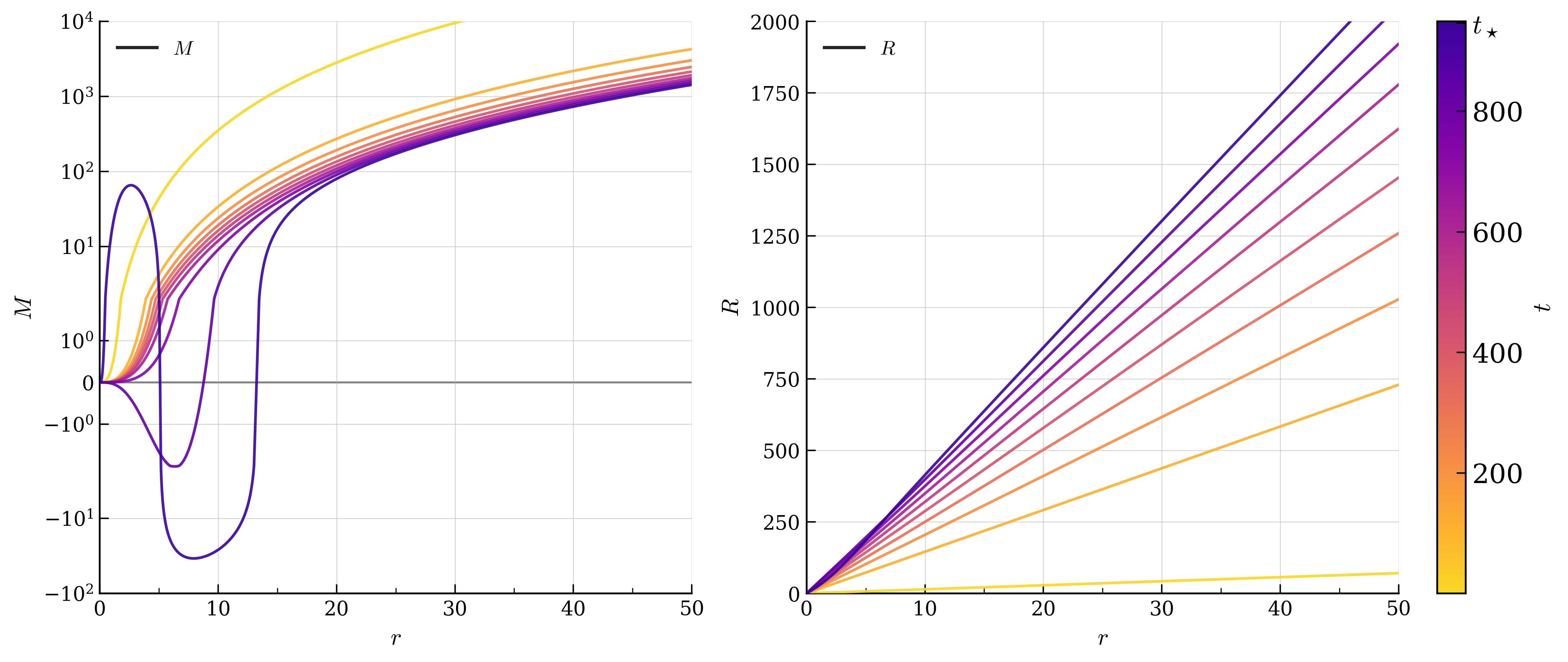}
    \caption{\footnotesize{\textbf{Left}: Misner-Sharp mass on a sequence of time slices approaching the first apparent horizon formation time $t_\star$. \textbf{Right}: Areal radius evolution}}
    \label{fig:MSmass}
\end{figure}
Figure~\ref{fig:rhotandphi} shows the evolution of the total local energy density $\rho_T(r)$ together with the Higgs profile $h(r)$. At early times the total density remains close to the homogeneous FRW background, while the Higgs field is described by a smooth superhorizon profile centred at the origin. As the evolution proceeds, the central Higgs field rolls deeper into the unstable region and the profile steepens near the core, while the outer region remains close to its false vacuum FRW background value. At the same time, the total density develops an increasingly localised central enhancement. The first collapse therefore produces a single marginally trapped surface enclosing the collapsing central negative potential region and the complete engulfment of the remaining exterior $V < 0$ region occurs subsequently and is quantified in Sec. IV.

Outside the central compact core the quantity $2M/R$ becomes negative over part of the profile. The left panel of Fig.~\ref{fig:MSmass} displays the Misner-Sharp mass during the initial collapse. Since $M$ is the Misner-Sharp quasi-local mass, this reflects the fact that the enclosed negative scalar potential contribution outweighs the positive kinetic, gradient and radiation terms in that region. Thus the Higgs patch is surrounded by a radial interval with negative enclosed Misner-Sharp mass, even though the central core still becomes sufficiently compact to satisfy $2M/R=1$ and form the first apparent horizon. The right panel shows that the areal radius $R$ remains a monotonically increasing function of $r$, although measurable deviations from FRW are present.  

\section{Late-time evolution: Supercritical and Subcritical}\label{section:LT}
Having established that the first nonlinear stage is common to all of the black hole forming configurations studied here, we now turn to the subsequent evolution of the positive potential barrier that survives outside the first apparent horizon. The diagnostics used to organise the subcritical and supercritical branches were introduced in Sec.~II C, while the representative initial parameters were specified in Sec.~II D. Here we show explicitly how the two late-time geometries arise in the simulations. In both cases the negative potential region has already undergone the collapse described in Sec.~III and formed the first primordial black hole. The distinction therefore arises from the subsequent evolution of the positive potential barrier outside this horizon.

\subsection{Supercritical evolution}

We first consider a representative supercritical configuration with
initial barrier compactness
\begin{equation}
    \frac{R_w}{\ell_{\sigma, \rm eff}}\simeq1.12,
\end{equation}
corresponding to $H_i/M_{ \rm Pl}=10^{-1}$ under the parametrisation of Sec.~II B. The barrier therefore already possesses substantial self-gravity on the initial hypersurface. Its potential energy is initially subdominant to the radiation,
\begin{equation}
    \frac{V_{\max}}{\rho_{ \rm fluid}(t_0)}
    \simeq 2.7\times10^{-5},
\end{equation}
but, as we show below, the surviving barrier subsequently evolves into a region in which
\begin{equation}
    \frac{V[h(t,r)]}{\rho_{ \rm fluid}(t,r)}\gtrsim1.
\end{equation}
The supercritical evolution therefore realises both features identified in Sec.~\ref{sec:branchdefinitions}. The first being the substantial barrier self-gravity and the second being the later development of a locally potential dominated region. The important point is that this secondary evolution begins only after the negative potential region has formed the first primordial black hole. We first describe the resulting spacetime geometry and then identify the stress-energy responsible for the transition.

\subsubsection{Geometric Spacetime Structure}\label{ss:GSS}

\begin{figure}[h!]
    \centering
    \includegraphics[width=0.8\linewidth]{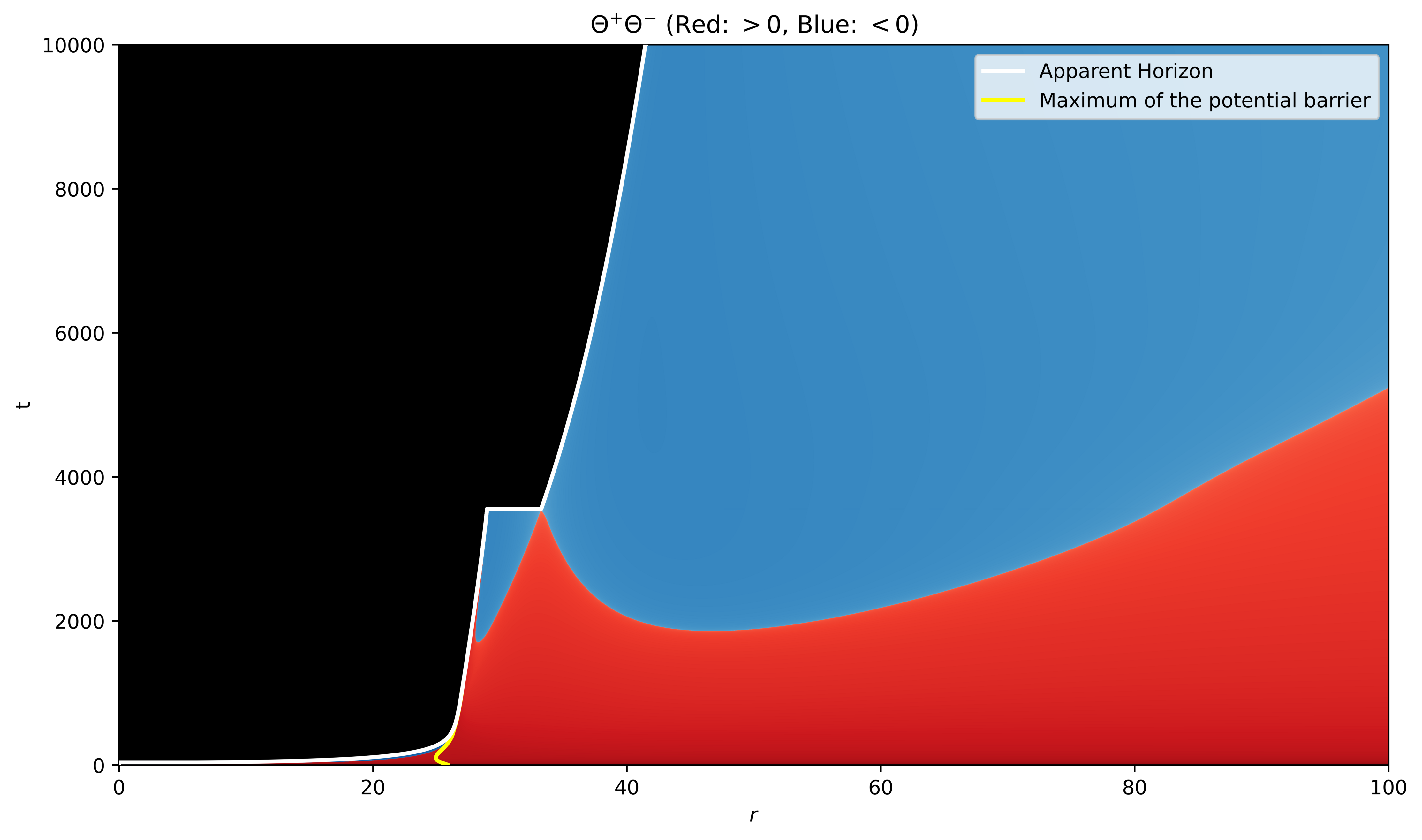}
    \caption{\footnotesize{Spacetime trapping structure of the supercritical Higgs barrier evolution in the $(r,t)$ plane. The colour map shows the sign of $\Theta^+\Theta^-$, with blue denoting the normal exterior region $(\Theta^+>0,\Theta^-<0)$ and red denoting regions where the two expansions have the same sign. The black region marks the excised black hole interior, the white curve traces the apparent horizon$(\Theta^+=0,\Theta^-<0)$ and the yellow curve follows the maximum of the Higgs potential barrier. The apparent horizon appears close to t = 0 on the y-axis since the collapse of the initial black hole occurs relatively quickly compared to the throat formation. In this simulation $t_\star=150$. The outward excursion of the apparent horizon indicates the formation of a secondary horizon branch during the transient wormhole phase.}}
    \label{fig:NGnew}
\end{figure}

Figure~\ref{fig:NGnew} shows the evolution of the radial null expansions as a function of radius and time. As the evolution proceeds, the maximum of the potential barrier, $h(r, t) = h_{ \rm max}$ travels outward, tracking the expansion of the barrier with the cosmological expansion. The maximum of the barrier remains outside the apparent horizon region throughout the evolution of the initial black hole, confirming that the barrier propagates through the causal exterior and is not immediately absorbed by the black hole. The distinguishing feature of the supercritical branch is the emergence of a secondary outer apparent horizon branch at $t\sim3500$. Before this transition, the apparent horizon traces the original black hole formed by the collapsing negative potential region. As the surviving barrier becomes sufficiently self-gravitating, a second trapped surface forms at larger radius, marking the onset of the non-monotonic wormhole-like geometry.
\begin{figure}[h!]
    \centering
    \includegraphics[width=0.8\linewidth]{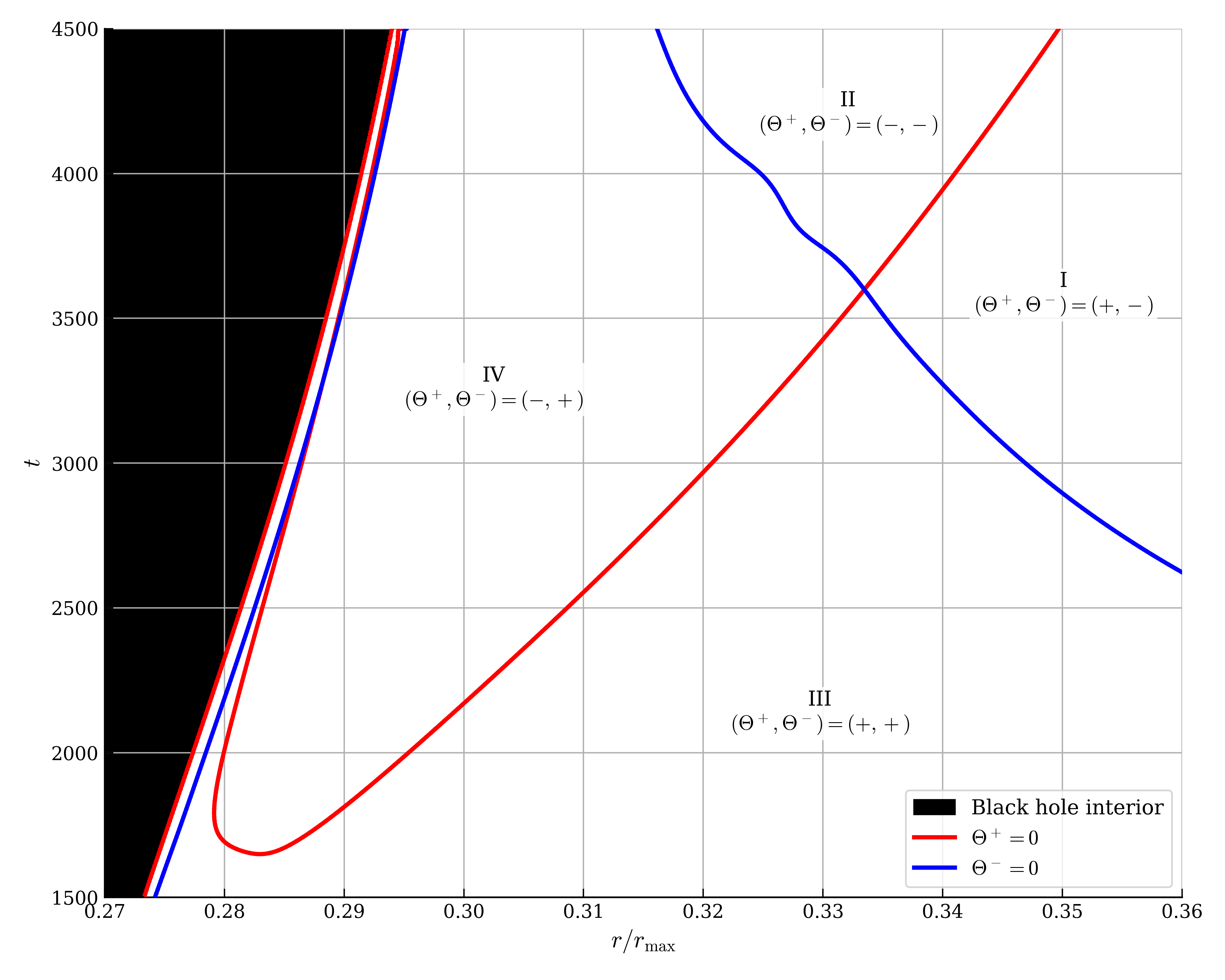}
    \caption{\footnotesize{Zoomed map of the null expansion structure in the throat forming region. The red curve denotes the $\Theta^+=0$ marginal surface and the blue curve denotes the $\Theta^-=0$ marginal surface. The labelled regions correspond to the normal exterior $(+,-)$, trapped black hole interior $(-,-)$, anti trapped FRW region $(+,+)$ and transient wormhole throat region $(-,+)$. The intersection of the two marginal surfaces marks the bifurcating horizon where $\Theta^+=\Theta^-=0$.}}
    \label{fig:NGnew_zoomedin}
\end{figure}

The local structure of the supercritical transition is shown in
Fig.~\ref{fig:NGnew_zoomedin}. The red and blue curves denote the marginal surfaces $\Theta^+=0$ and $\Theta^-=0$, respectively. Before the transition, the spacetime consists of the trapped black hole interior, the normal exterior (region I) and the anti-trapped FRW background (region III). During the transition, a fourth region appears between the two marginal surfaces with
\begin{equation}
    (\Theta^+,\Theta^-)=(-,+),
\end{equation}
so that outgoing null rays converge while ingoing null rays diverge. This is the causal signature of the transient wormhole throat. The two marginal surfaces intersect at a \emph{bifurcating trapping horizon}. Unlike an ordinary black hole apparent horizon, for which $\Theta^+=0$ while $\Theta^-<0$, a bifurcating horizon is a marginal surface on which both future-directed null expansions vanish simultaneously,
\begin{equation}
    \Theta^+=\Theta^-=0.
\end{equation}
Such surfaces arise naturally in trapping horizon descriptions of strongly curved spacetimes and have also been identified in the Type-B branch of primordial black hole formation
\cite{Hayward_1996,Kopp2011,Uehara_2025,
shimada2024primordialblackholeformation}.
Using
\begin{equation}
    \Theta_\pm=\frac{\sqrt{2}}{R}(U\pm\Gamma),
\end{equation}
the bifurcation condition is equivalent to
\begin{equation}
    U=\Gamma=0.
\end{equation}

Region IV exists only for a finite interval and is bounded by two
intersections of the marginal surfaces. The first marks the appearance of the throat region, while the second marks the end of this transient bifurcating structure. The resolved throat and bifurcating trapping horizon are consistent with the formation of a child-universe branch that becomes causally disconnected from the parent exterior, while the parent spacetime remains connected only to the outer black hole horizon.

There is an important distinction between the present evolution and the usual discussion of Type-B PBH formation. In the PBH literature, Type-I and Type-II refer to the geometry of the initial fluctuation, whereas Type-A and Type-B refer to the trapping horizon structure that forms dynamically \cite{Kopp2011,Uehara_2025,shimada2024primordialblackholeformation}. A Type-II initial fluctuation possesses a stationary point of the areal radius on the initial hypersurface, while the standard Type-I configuration has a monotonic areal radius. By contrast, a Type-B PBH is characterised by the appearance of a bifurcating trapping horizon. Our initial data are manifestly Type-I in this sense. On the initial hypersurface,
\begin{equation}
    R(t_0,r)=r,
    \qquad
    \Gamma(t_0,r)=1,
\end{equation}
so the areal radius is everywhere monotonic and no throat or stationary point is present. Indeed, the first collapse produces the ordinary Type-A horizon described in Sec.~III. The later evolution nevertheless generates the bifurcating horizon characteristic of the Type-B branch. The sequence realised here is therefore
\begin{equation}
    \text{Type-I initial geometry}
    \;\longrightarrow\;
    \text{Type-A first horizon}
    \;\longrightarrow\;
    \text{Type-B horizon structure}.
\end{equation}
The transition from the Type-A to the Type-B geometry is produced
dynamically by the surviving positive potential barrier.

\begin{figure}[h!]
    \centering
    \includegraphics[width=0.8\linewidth]{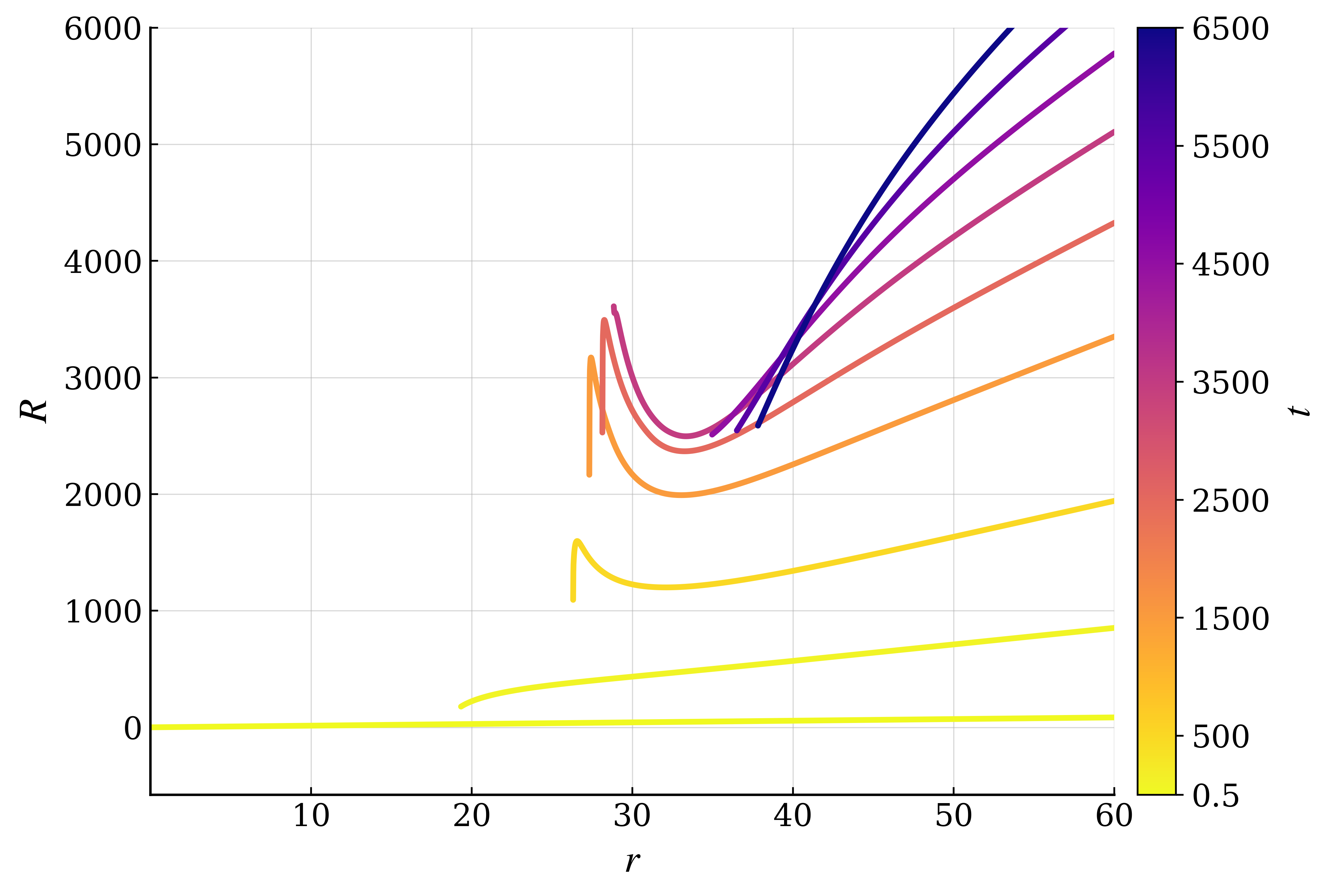}
    \caption{\footnotesize{Areal radius profiles $R(r)$ on constant time slices during the supercritical evolution. At early times $R(r)$ is monotonic, while during the wormhole phase it develops a local minimum, identifying the throat where $\partial R/\partial r=0$, equivalently $\Gamma=0$. The non-monotonic profiles show the formation and subsequent evolution of the wormhole-like geometry.}}
    \label{fig:arealradius}
\end{figure}

The areal radius profiles in Fig.~\ref{fig:arealradius} provide an independent geometric confirmation of the throat interpretation. Before the supercritical transition, $R(r,t)$ is monotonic on constant-time slices. During the throat phase it develops a local minimum satisfying
\begin{equation}
    \frac{\partial R}{\partial r}=0,
    \qquad
    \Gamma=0,
\end{equation}
which identifies the geometric neck of the wormhole. Region IV therefore corresponds to a genuine non-monotonicity of the spatial geometry. 

\begin{figure}[h!]
    \centering
    \includegraphics[width=0.8\linewidth]{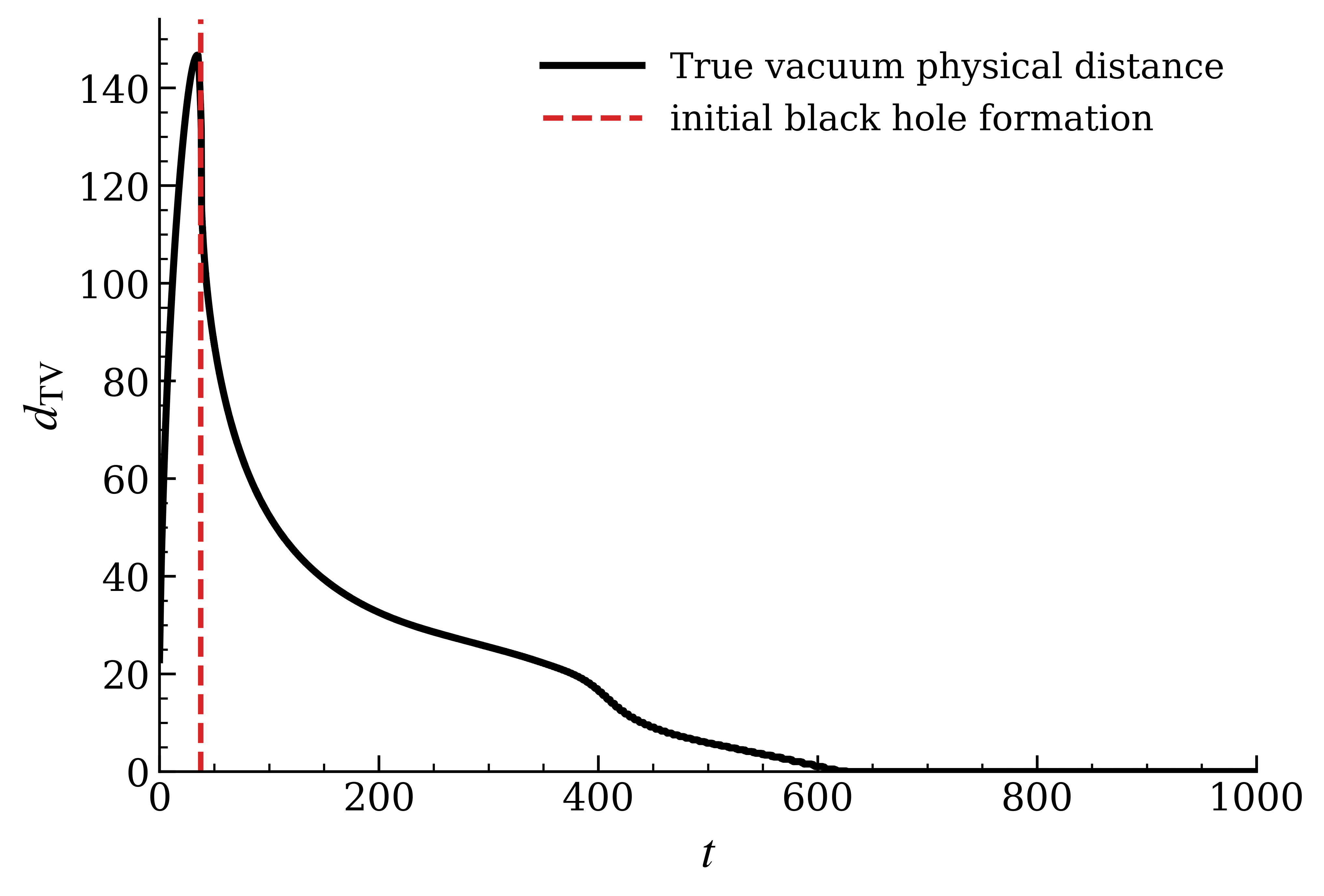}
    \caption{\footnotesize{Evolution of the proper radial extent $d_{\rm TV}$ of the  $V<0$ region in the representative supercritical evolution. The red dashed line marks the
    formation of the first apparent horizon.}}
    \label{fig:physdist}
\end{figure}

Having established the throat geometry, we now ask whether any part of the $V<0$ region remains exposed to the parent FRW universe during this later transition. To quantify this, we define the proper radial extent of the negative potential region by
\begin{equation}
d_{\rm TV}(t)=
\begin{cases}
\displaystyle
\int_{0}^{r_{V=0}(t)} B(r,t)\,dr,
& t<t_\star,
\\[1.2ex]
\displaystyle
\int_{r_{\rm AH}(t)}^{r_{V=0}(t)} B(r,t)\,dr,
& t\geq t_\star
\quad \text{and} \quad
r_{\rm AH}(t)<r_{V=0}(t),
\\[1.2ex]
0,
& t\geq t_\star
\quad \text{and} \quad
r_{\rm AH}(t)\geq r_{V=0}(t).
\end{cases}
\label{eq:dTV}
\end{equation}
where $B(r,t)\,dr$ is the proper radial line element on a constant-time slice, $r_{V=0}(t)$ denotes the outward $V=0$ crossing that bounds the central connected $V<0$ region, and $r_{\rm AH}(t)$ is the coordinate position of the black hole apparent horizon. Before the first apparent horizon forms, $d_{\rm TV}$ therefore measures the proper radial size of the full negative potential region. After horizon formation, it measures only the portion of that region remaining outside the apparent horizon. We set $d_{\rm TV}=0$ once the apparent horizon has engulfed the $V=0$ surface.

Immediately after the first apparent horizon forms, a finite negative potential region remains outside it and $d_{\rm TV}$ reaches a maximum as shown in Fig.~\ref{fig:physdist}. The separation then decreases to zero at $t\sim650$, showing that the original horizon completely encloses the $V<0$ region. This occurs before the transient wormhole phase that develops at $t\sim1500$. The later supercritical transition therefore does not correspond to an exposed negative potential region expanding into the parent universe.

\subsubsection{What drives the secondary transition?}\label{sec:drivesthroat}

Having established the formation of the secondary horizon and wormhole throat geometrically, we now identify the stress-energy responsible for this transition. The negative potential region cannot be the source, since it is already enclosed by the first black hole before the secondary transition occurs. Instead, the relevant matter component is the surviving positive potential
barrier.

\begin{figure}[h!]
    \centering
    \includegraphics[width=0.8\linewidth]{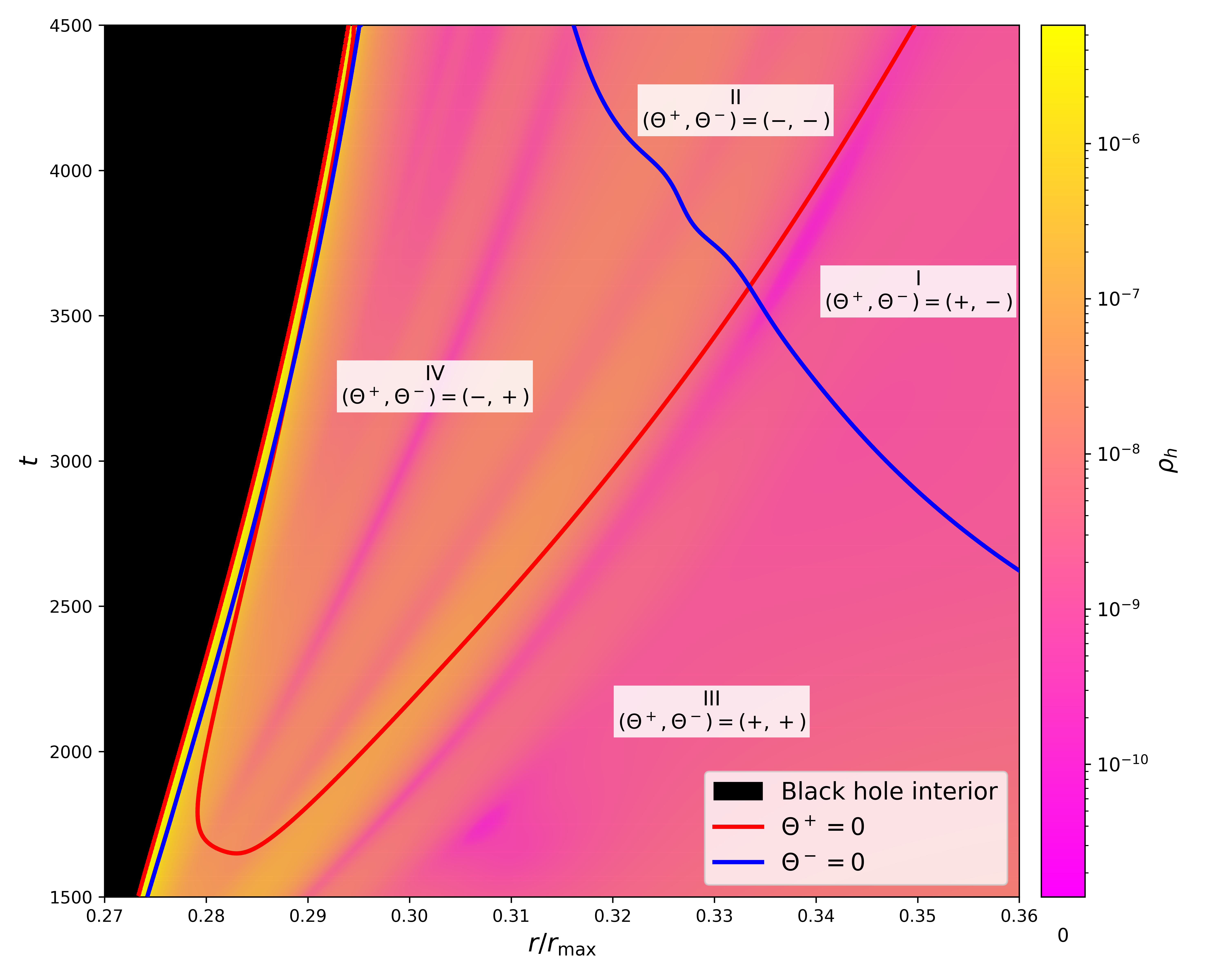}
    \caption{\footnotesize{{The expansion of the null geodesics structure of the supercritical evolution overlaid with the scalar field energy density $\rho_{h}$ (background colour). The red and
    blue curves are the $\Theta^{+} = 0$ and $\Theta^{-} = 0$ loci respectively. This simulation was allowed to run beyond the time of the bifurcating horizon.}}}
    \label{fig:signsrhosf}
\end{figure}
Figure~\ref{fig:signsrhosf} shows the scalar-field energy density overlaid on the geodesic expansion structure of Fig.~\ref{fig:NGnew_zoomedin}. After the first apparent horizon forms, the remaining scalar energy becomes concentrated in
a band outside the black hole that follows the positive potential barrier. This band subsequently overlaps the region in which the secondary horizon branch and throat develop, directly associating the transition with the surviving barrier rather than the collapsed negative potential region.

When the positive potential barrier becomes locally dominant over the radiation background, its stress-energy becomes sufficiently vacuum energy-like to support accelerated expansion relative to the surrounding radiation dominated universe. As the potential becomes increasingly dominant over the kinetic and radiation contributions, the local stress-energy approaches $p\simeq-\rho$. Local potential domination alone is not sufficient as the inflating region must also be sufficiently extended and self-gravitating. When its characteristic size becomes comparable to its gravitational length, the differential expansion between the inner barrier dominated region and the parent FRW exterior drives the areal radius toward a non-monotonic profile, producing a throat and the
Type-B/child-universe branch. The stress-energy diagnostics below provide the GR description of this local inflation picture.

To connect this local inflation picture directly to the Einstein equations, we consider the radial stress $T^1{}_1$ and the active gravitational source
\begin{equation}
    S \equiv 4\pi\left(T^0_{\ 0}+T^1_{\ 1}+2T^2_{\ 2}\right).
\end{equation}The radial stress enters the evolution of the areal radius velocity through
\begin{equation}
    \dot U = -\frac{1-\Gamma^2+U^2}{2R} -4\pi R T^1_{\ 1},
    \label{eq:Udot_source}
\end{equation}
so that $T^1_{\ 1}<0$ provides a positive contribution to $\dot U$ and drives the areal radius outwards. Equation~\eqref{eq:Udot_source} takes this particularly transparent form in the geodesic slicing $A=1$ adopted here. In a more general slicing, lapse-gradient terms also
contribute to the coordinate acceleration. The physical interpretation does not rely on this decomposition alone as the vacuum-like stress of the barrier is accompanied by geometric signatures as we discussed above.

Similarly, $S$ enters the evolution of the trace of the extrinsic curvature,
\begin{equation}
    \dot K  = -\left(K-\frac{2U}{R}\right)^2 -2\left(\frac{U}{R}\right)^2 -S .
    \label{eq:Kdot_source}
\end{equation}
 Negative $S$ gives a positive contribution to $\dot K$ and therefore supports local accelerated expansion around the barrier. For the scalar fluid system,
\begin{equation}\label{eq:S_scalarfluid}
    S  = 4\pi \left[2\dot h^2 - 2V(h) +  \rho_{ \rm fluid}
        \left(\frac{(1+w)(1+v^2)}{1-v^2} + 2w \right)\right].
\end{equation}
The scalar gradient contribution cancels from $S$, because the radial gradient contributes positively to the radial pressure but negatively to the angular pressures. Consequently,
$S<0$ requires the positive potential contribution to dominate over the kinetic and radiation terms, so that the barrier behaves locally like a positive vacuum-energy source, producing gravitational defocusing rather than focusing. 

This also distinguishes the two stages of the evolution. In the
negative potential region,
\begin{equation}
    T^1_{\ 1}\sim -V>0,
    \qquad
    S\sim -2V>0,
\end{equation}
so the negative potential does not support local inflation and instead contributes to the initial collapse. By contrast, in the surviving positive potential barrier the potential can become sufficiently dominant that
\begin{equation}
    T^1_{\ 1}<0, \qquad S<0,
\end{equation}
providing the outward acceleration and the stress-energy signature of the locally inflating region associated with the secondary transition.

\begin{figure}[h!]
    \centering
    \includegraphics[width=0.9\linewidth]{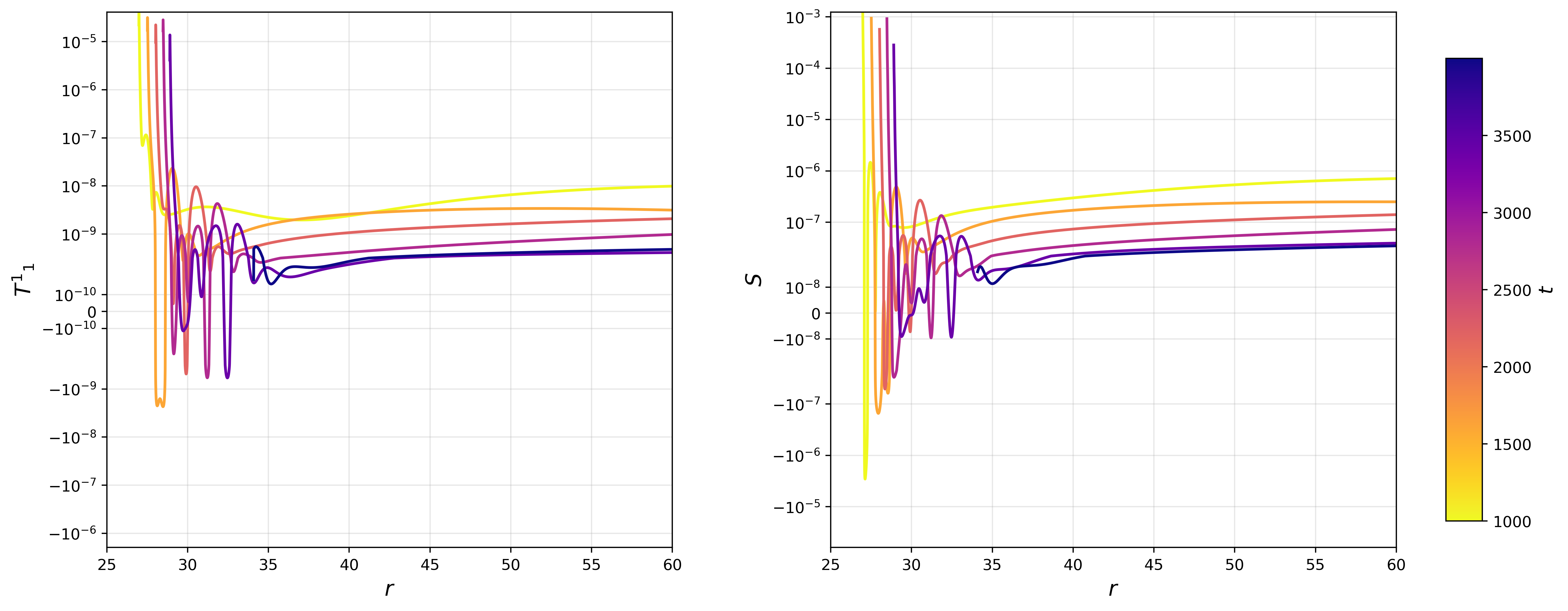}
    \caption{\footnotesize{{Radial profiles of the radial stress $T^1{}_1$ and the
    active gravitational source $S=4\pi\big(T_{00}+T^1{}_1+2T^2{}_2\big)$ on a sequence of time
    slices during the supercritical evolution.}}}
    \label{fig:ST11}
\end{figure}

Figure~\ref{fig:ST11} confirms this picture. As the secondary horizon branch develops, both $T^1_{\ 1}$ and $S$ become negative in the surviving barrier region. The same region therefore provides both the negative radial stress that accelerates the areal radius outwards and the negative active gravitational
source characteristic of the locally inflating barrier required to form the throat. Together with the bifurcating trapping horizon and the local minimum of $R$, this identifies the positive potential barrier as the driver of the transient inflating child-universe geometry.

\begin{figure}[h!]
    \centering
    \includegraphics[width=0.8\linewidth]{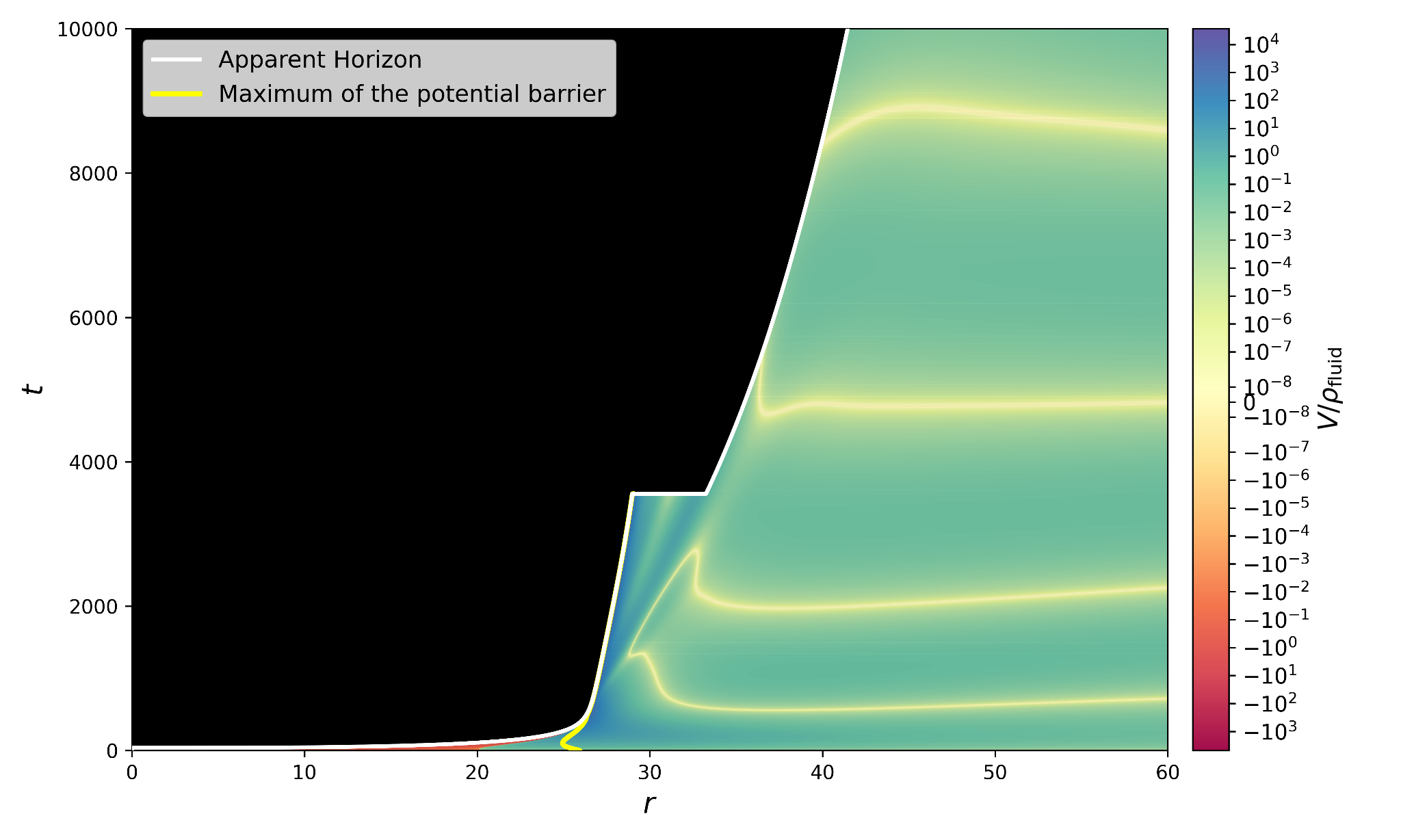}
    \caption{\footnotesize{{Ratio $V/\rho_{\rm fluid}$ during the supercritical evolution. A positive barrier potential dominated region with $V/\rho_{\rm fluid}>1$ develops outside the original apparent horizon and tracks the region where the secondary horizon branch forms.}}}
    \label{fig:Voverrho}
\end{figure}

The connection to the local inflation picture is shown explicitly in Fig.~\ref{fig:Voverrho}, where we plot $V/\rho_{\rm fluid}$. A region with
\begin{equation}
    \frac{V}{\rho_{\rm fluid}}>1
\end{equation}
develops outside the original apparent horizon and closely follows the secondary $\Theta^+=0$ branch. Crucially, this ratio is positive as the region is dominated by the surviving positive potential barrier rather than by the $V<0$ part of the scalar configuration.

\begin{figure}[h!]
    \centering
    \includegraphics[width=0.8\linewidth]{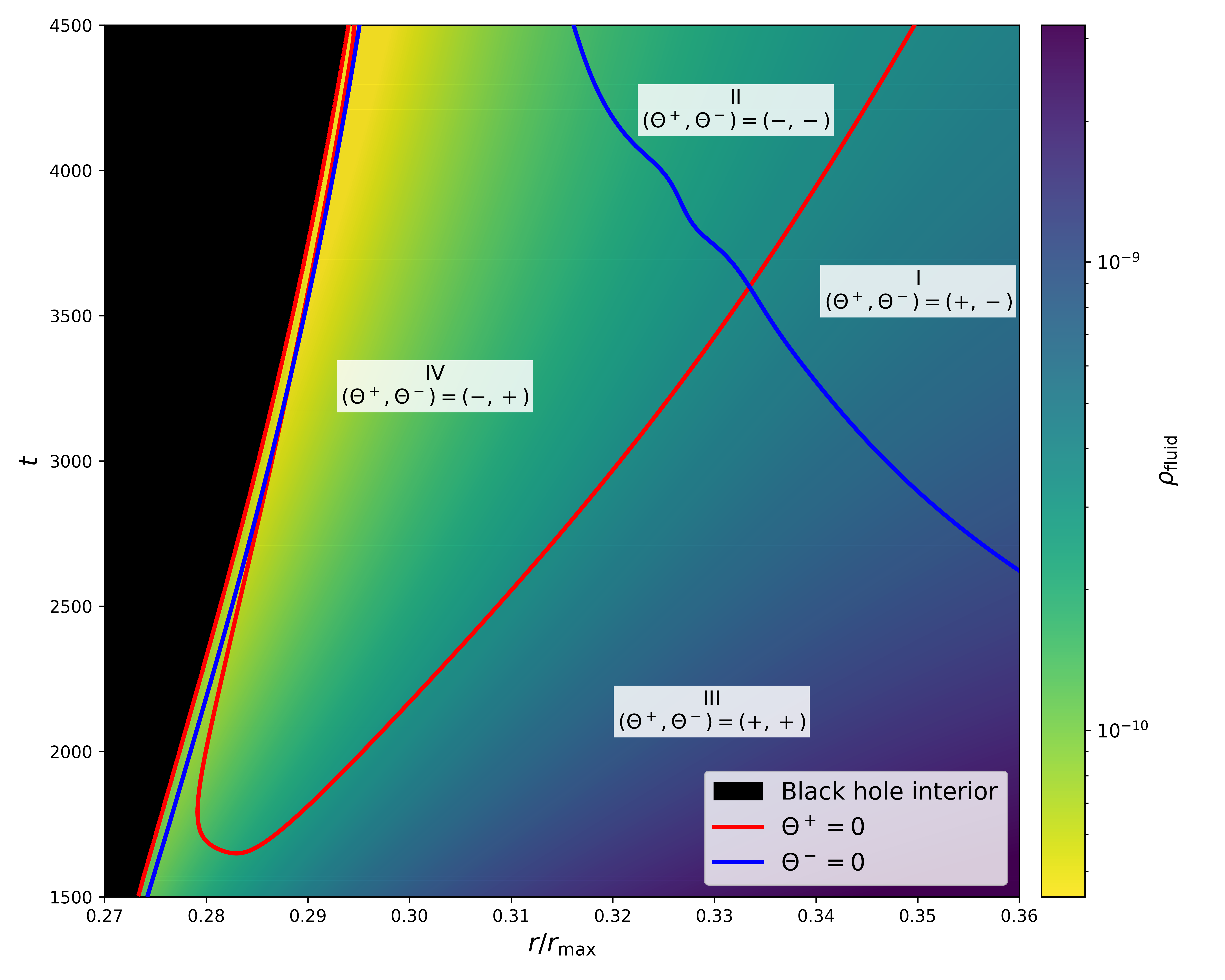}
    \caption{\footnotesize{{The expansion of the null geodesic structure of the supercritical evolution overlaid with the scalar field energy density $\rho_{ \rm fluid}$ (background colour). The red and blue curves are the $\Theta^{+} = 0$ and $\Theta^{-} = 0$ loci respectively. This simulation was allowed to run beyond the time of the bifurcating horizon.}}}
    \label{fig:signsrhopf}
\end{figure}
The complementary radiation fluid evolution is shown in
Fig.~\ref{fig:signsrhopf}. In contrast to the localised scalar structure, $\rho_{\rm fluid}$ remains comparatively smooth across the secondary transition and does not track the throat forming region. The radiation responds to the evolving geometry but does not provide the localised source of the transition. Together with Figs.~\ref{fig:ST11} and~\ref{fig:Voverrho}, this confirms that the secondary geometry is driven by the locally potential dominated, self-gravitating scalar barrier.

\subsection{Subcritical Evolution}
We next consider the representative subcritical configuration used in
Sec.~III to demonstrate the common first collapse. For this
configuration,
\begin{equation}
    \left.
    \frac{R_w}{\ell_{\sigma, \rm eff}}
    \right|_{t_0}
    \simeq0.02 .
\end{equation}
The distinction from the supercritical example therefore lies in the subsequent evolution of the positive potential barrier. As we show below, while the barrier remains outside the original apparent horizon it does not become locally dominant over the radiation and does not generate the secondary geometric transition.
The original apparent horizon grows smoothly outwards and progressively engulfs the remaining scalar configuration.

\subsubsection{Geometric Spacetime Structure}

\begin{figure}[h!]
    \centering
    \includegraphics[width=0.8\linewidth]{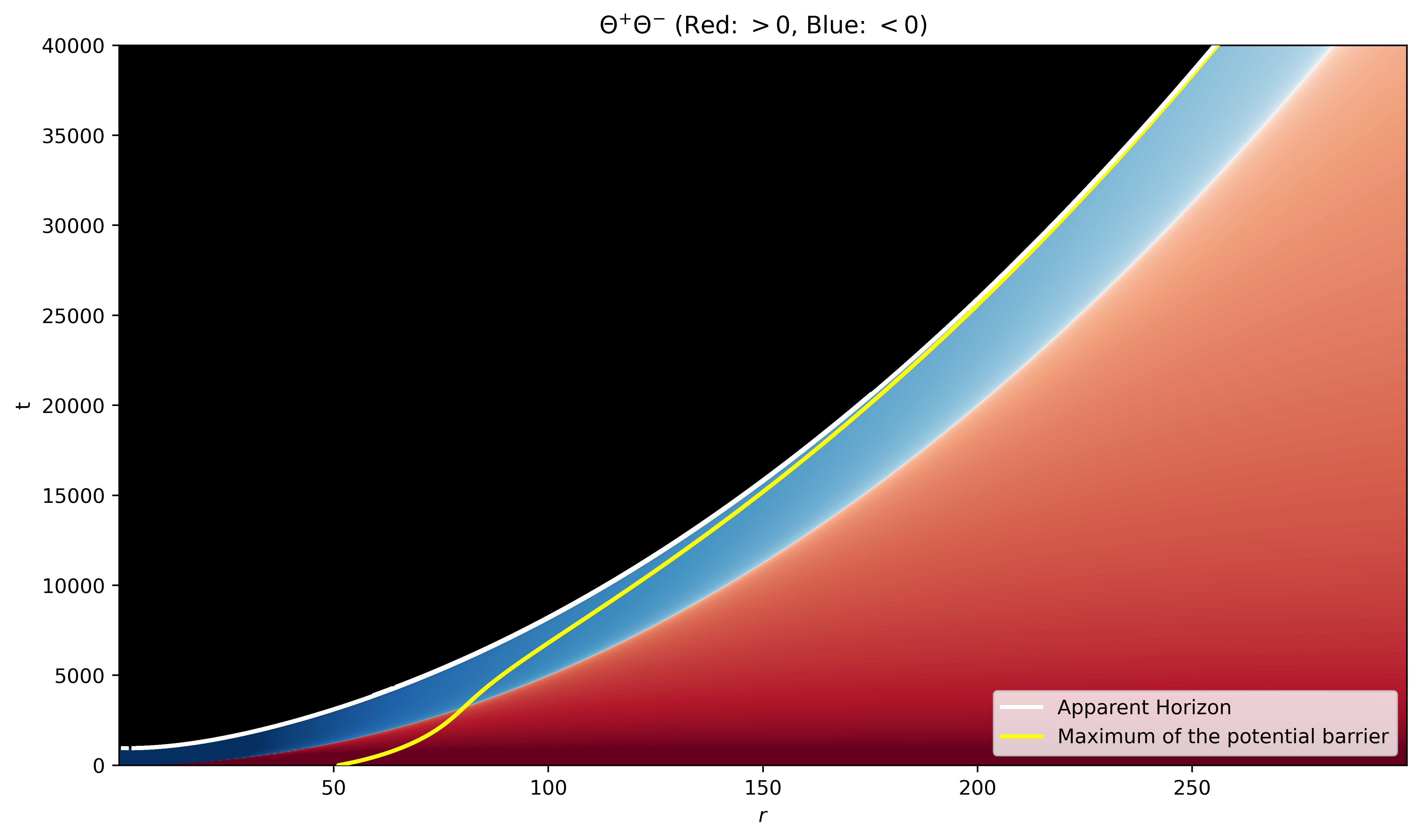}
    \caption{\footnotesize{Spacetime trapping structure of the subcritical Higgs barrier evolution for $R_w/\ell_{\sigma, \rm eff} \simeq 0.02$. The colour map shows the sign of $\Theta^+\Theta^-$, with blue denoting the normal exterior region $(\Theta^+>0,\Theta^-<0)$ and red denoting regions where the two expansions have the same sign. The black region marks the excised black hole interior, the white curve traces the apparent horizon and the yellow curve follows the maximum of the Higgs potential barrier.}}
    \label{fig:NG_-15}
\end{figure}

Figure~\ref{fig:NG_-15} shows the trapping structure using the same
conventions as Fig.~\ref{fig:NGnew}, allowing a direct comparison with the supercritical evolution. The key difference is the absence of any secondary horizon structure. After the first black hole forms, the apparent horizon remains a single continuous branch and grows smoothly outwards through the surrounding scalar configuration. The yellow curve follows the positive potential barrier, $h(r,t)=h_{\max}$. The barrier remains outside the apparent horizon for a finite time and passes through the normal exterior region, $(\Theta^+,\Theta^-)=(+,-)$. As in the supercritical case, the first horizon forms inside the $V<0$ region rather than enclosing it completely, so part of the negative potential region initially remains outside the black hole. In the subcritical evolution, however, this exterior remnant progressively shrinks and is eventually engulfed by the growing apparent horizon. No second $\Theta^+=0$ branch, bifurcating horizon or throat develops. 

In contrast to the supercritical evolution, the surviving positive potential barrier does not develop an extended locally inflating region capable of reorganising the exterior geometry. Although the scalar configuration perturbs the spacetime away from the FRW background, it does not generate the secondary throat transition.

\begin{figure}[h!]
    \centering
    \includegraphics[width=0.8\linewidth]{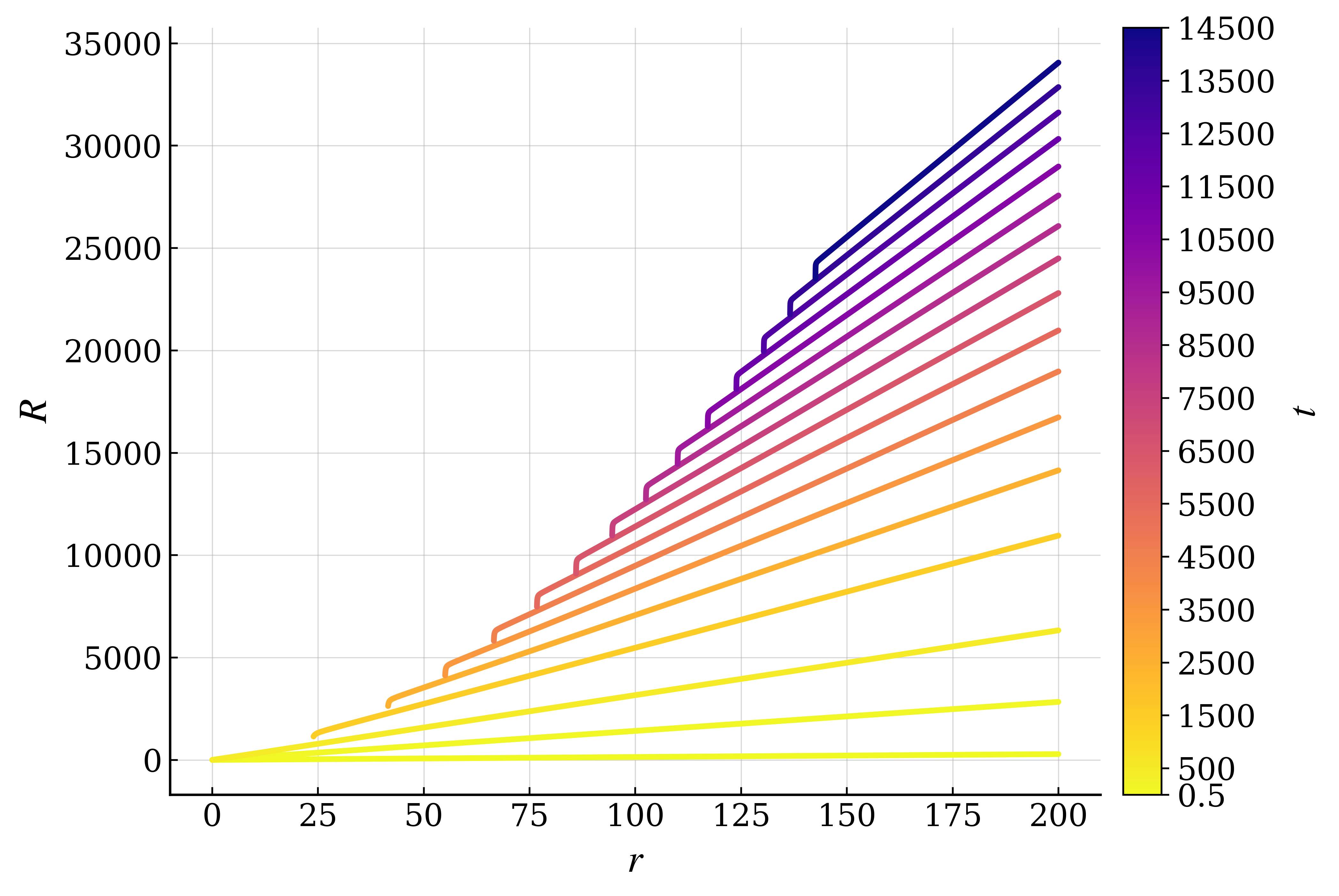}
    \caption{\footnotesize{Areal radius profiles $R(r)$ on constant-time slices during the subcritical evolution. Although the Higgs configuration deforms the geometry away from pure FRW, $R(r)$ remains monotonic on all slices shown. The absence of a local minimum, equivalently the absence of $\partial R/\partial r=0$ or $\Gamma=0$ outside the original horizon, shows that the subcritical barrier does not
generate a wormhole throat.}}
    \label{fig:R-15}
\end{figure}

The absence of a throat is confirmed directly by the areal radius profiles in Fig.~\ref{fig:R-15}. Although the exterior geometry is visibly perturbed, $R(r,t)$ remains monotonic on every slice shown and there is no point outside the original horizon satisfying
\begin{equation}
    \frac{\partial R}{\partial r}=0, \qquad
    \Gamma=0 .
\end{equation}
This provides a direct geometric distinction from the supercritical branch, where the corresponding local minimum of $R$ forms the neck of the transient wormhole. The subcritical spacetime instead retains a single connected black hole exterior embedded in the surrounding radiation dominated FRW universe.

\begin{figure}[h!]
    \centering
    \includegraphics[width=0.7\linewidth]{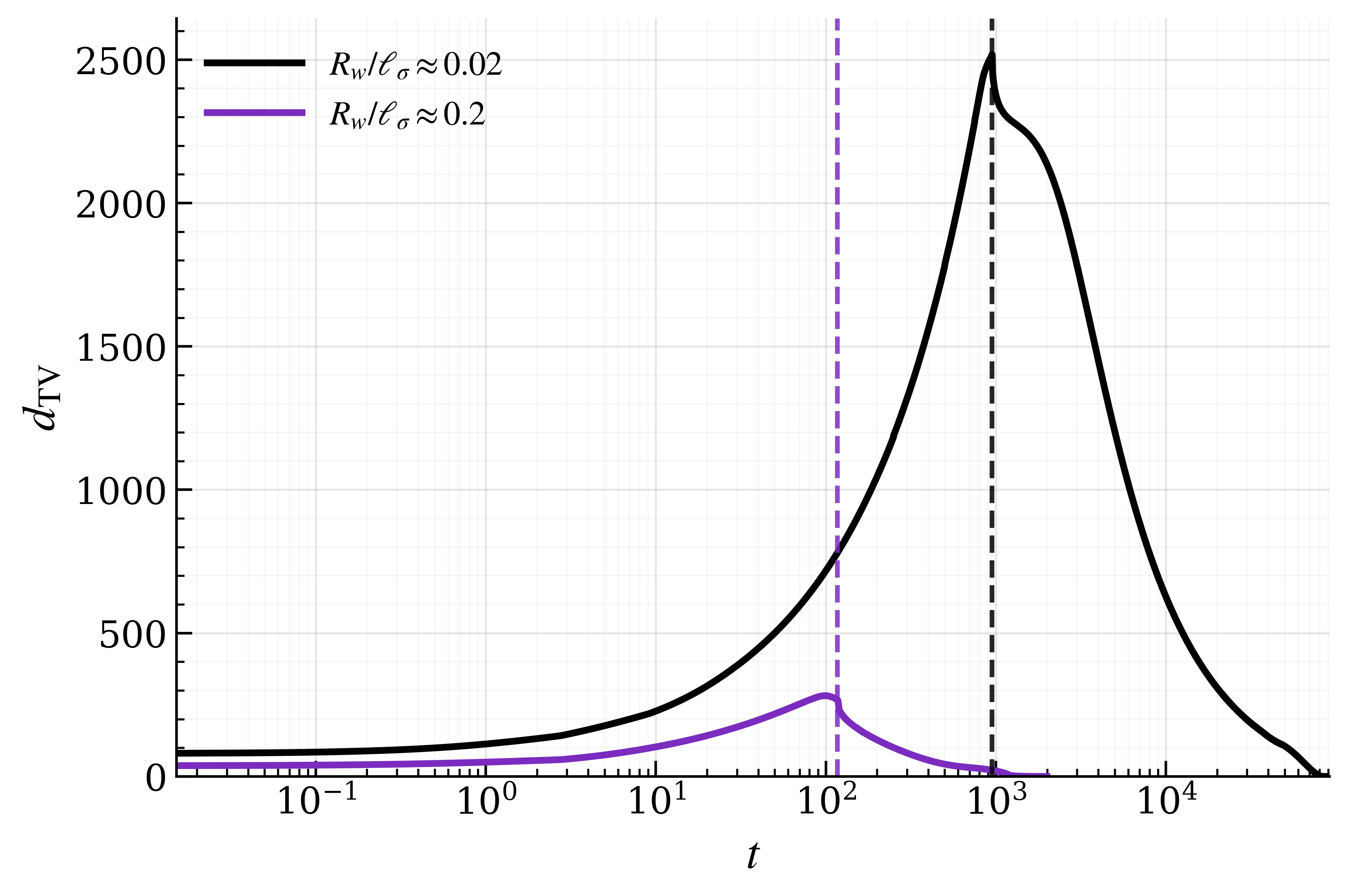}
    \caption{\footnotesize{Evolution of the proper radial extent $d_{\rm TV}$ of the $V<0$ region for two subcritical evolutions with initial $R_w/\ell_{\sigma, \rm eff}\simeq0.02$ (black) and $R_w/\ell_{\sigma,\rm eff}\simeq0.2$ (purple). The vertical dashed lines mark the formation of the first apparent horizon in each simulation. In both cases $d_{\rm TV}$ subsequently decreases to zero as the apparent horizon grows outward and completely engulfs the remaining exterior $V<0$ region.}}
    \label{fig:dtvsubcritical}
\end{figure}

Figure~\ref{fig:dtvsubcritical} shows the evolution of $d_{\rm TV}$ for two subcritical configurations. The first is the representative simulation discussed above, with
\begin{equation}
    \frac{H_i}{M_{ \rm Pl}}=10^{-5},
    \qquad
    \frac{R_w}{\ell_{\sigma, \rm eff}}\simeq0.02 .
\end{equation}
For comparison, we also show a second simulation with the same dimensionless initial profile,
\begin{equation}
    \frac{h_{\max}}{H_i}=4,
    \qquad
    \frac{h_i}{h_{\max}}=1.5,
    \qquad
    H_i r_L=20,
\end{equation}
but with the Hubble scale increased to
\begin{equation}
    \frac{H_i}{M_{ \rm Pl}}\simeq10^{-4},
\end{equation}
for which the initial barrier compactness increases to
\begin{equation}
    \frac{R_w}{\ell_{\sigma, \rm eff}}\simeq0.2 .
\end{equation}

At the time of first apparent horizon formation, $d_{\rm TV}$ is finite in both simulations, confirming that the first black hole forms within the collapsing $V<0$ region rather than enclosing it completely. The separation subsequently decreases to zero as the apparent horizon grows outwards, with $d_{\rm TV}=0$ marking the complete engulfment of the remaining exterior negative potential
region. Although the engulfment time differs between the two configurations, both remain subcritical as no secondary horizon or throat forms and the original apparent horizon ultimately encloses the entire $V<0$ region.

\subsubsection{Scalar-Fluid Spacetime Structure}
The absence of the secondary geometric transition has a direct
stress-energy interpretation. In the subcritical branch, while the surviving positive potential barrier remains outside the apparent horizon and dynamically relevant to the exterior, it remains locally subdominant to the radiation. Therefore the locally inflating source required to form a throat does not occur.

\begin{figure}[h!]
    \centering
    \includegraphics[width=0.8\linewidth]{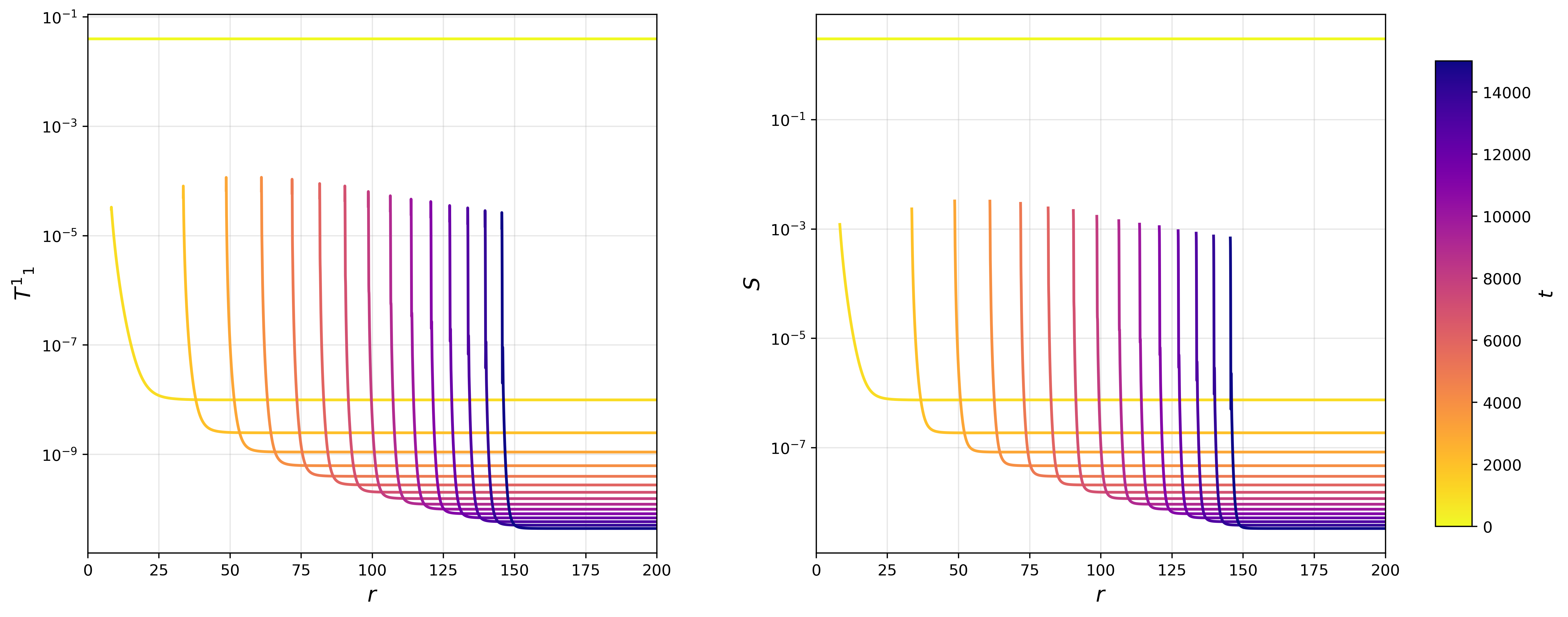}
    \caption{\footnotesize{Radial profiles of the radial stress $T^1{}_1$ and the active gravitational source $S$
on a sequence of time slices during the subcritical evolution. Both quantities remain positive at all radii and at all times shown. The Higgs barrier therefore never provides the repulsive gravitational source needed to drive a non-monotonic areal radius profile, explaining why no secondary horizon branch or wormhole throat forms in the subcritical regime.}}
    \label{fig:ST11_subcritical}
\end{figure}

Figure~\ref{fig:ST11_subcritical} shows the radial stress $T^1{}_1$ and active gravitational source $S$ and should be compared directly with the supercritical profiles in Fig.~\ref{fig:ST11}. Neither quantity becomes negative in the exterior barrier region. The stress-energy signature associated with the locally inflating
supercritical barrier is therefore absent. This is consistent with the monotonic areal radius
profiles in Fig.~\ref{fig:R-15}.

\begin{figure}[h!]
    \centering
    \includegraphics[width=0.75\linewidth]{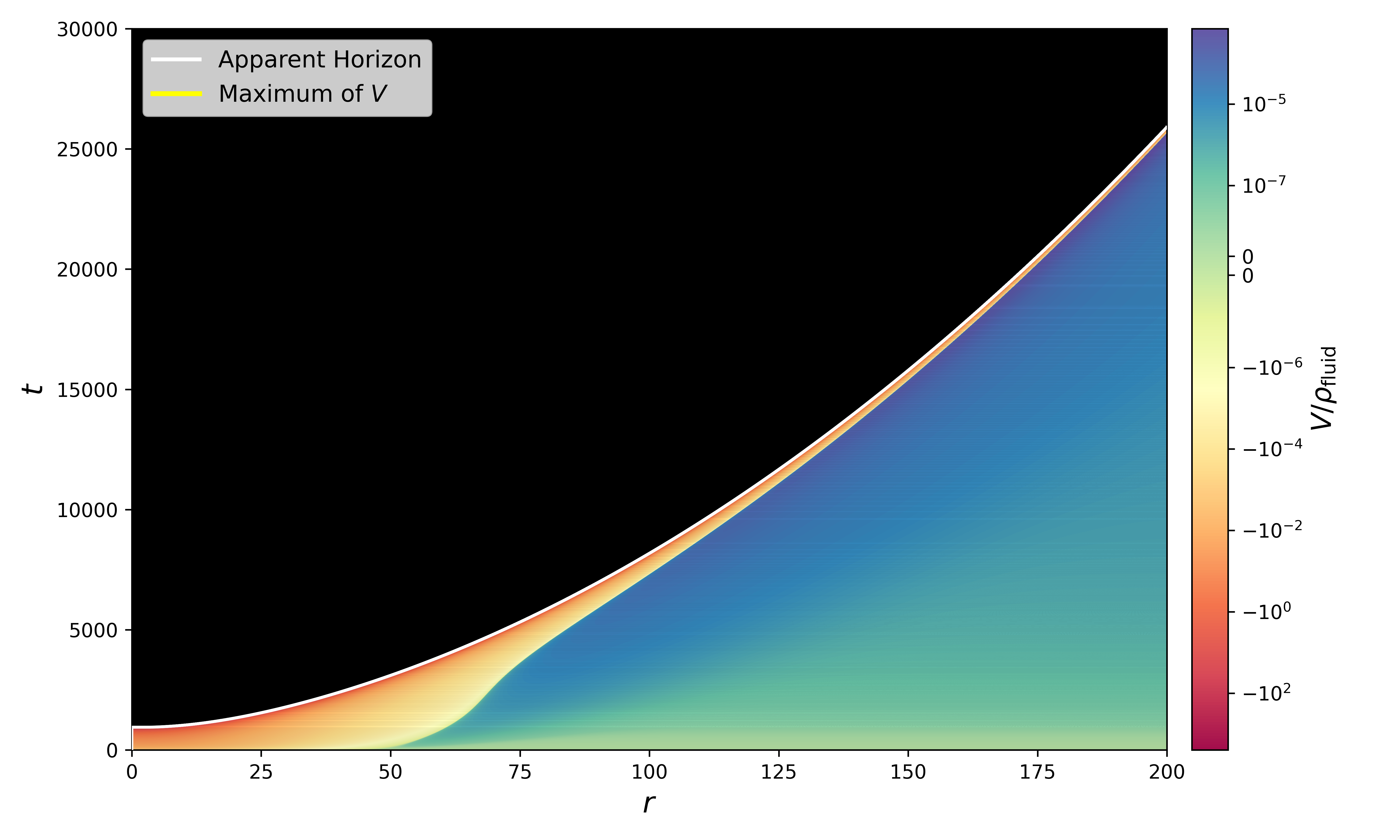}
    \caption{\footnotesize{Spacetime structure of the subcritical evolution overlaid with the ratio
$V/\rho_{ \rm fluid}$. The white curve marks the apparent horizon and the yellow curve tracks the maximum of the Higgs potential barrier. Regions with $V/\rho_{\rm fluid}<0$ correspond to the unstable $V<0$ part of the scalar configuration. The positive potential therefore never becomes locally dominant relative to the radiation outside the apparent horizon, providing the local counterpart to the weak barrier self-gravity discussed in the
text.}}
    \label{fig:Vrho_subcritical}
\end{figure}

The same conclusion is seen directly in Fig.~\ref{fig:Vrho_subcritical}. Immediately after the
first apparent horizon forms, part of the $V<0$ region remains outside the horizon, corresponding to the finite $d_{\rm TV}$ shown in Fig.~\ref{fig:dtvsubcritical}. Within this region $V/\rho_{\rm fluid}$ is negative and its magnitude can
exceed unity. This represents domination by the negative scalar potential and therefore does not correspond to a locally inflating region. The relevant comparison with the supercritical branch is instead the positive potential barrier. While this barrier remains outside the apparent horizon, it does not develop an extended region with
\begin{equation}
    \frac{V}{\rho_{\rm fluid}}>1.
\end{equation}
The locally inflating positive potential region seen in the supercritical case therefore does not develop during the dynamically relevant exterior phase. Together with the smaller initial barrier self-gravity, this explains why no secondary throat or bifurcating horizon forms.

We do not expect the inequality $V/\rho_{\rm fluid}<1$ to hold indefinitely at late times. The radiation density decreases through cosmological redshifting and can also be depleted locally by accretion onto the black hole, so a fixed scalar potential scale may eventually become large relative to $\rho_{\rm fluid}$. However, in the subcritical evolution the positive potential hilltop configuration has already been engulfed by the
growing apparent horizon before this can produce a locally inflating exterior region. Any subsequent decrease of the radiation density therefore does not alter the absence of the secondary exterior transition.

\section{Discussion}
We have presented fully nonlinear numerical simulations of the gravitational collapse of a Higgs-like scalar field in a radiation dominated FRW background, with initial conditions motivated by rare stochastic inflationary excursions beyond the maximum of the Higgs potential. The main results of our simulations are that the collapse of a metastable scalar fluctuation is controlled by two distinct pieces of the field configuration. The negative potential region determines whether the initial collapse occurs, whereas the positive potential barrier determines the subsequent global geometry. These two processes need not have the same outcome or occur on the same timescale. In particular, the formation of a
black hole around the runaway core does not imply that the remaining scalar configuration must simply accrete into it. If the surviving barrier becomes both locally important relative to the radiation and sufficiently self-gravitating, it can instead reorganise the exterior spacetime and generate a bifurcating trapping horizon and a transient child-universe
geometry.

This provides a useful way of interpreting the different
outcomes discussed in the Higgs-instability literature. The question is not whether an over-the-barrier region ``expands'' or ``collapses'' as a configuration can do both at different stages. The negative potential region can collapse first, while the positive potential part of the same fluctuation subsequently drives an expanding geometry. In the supercritical
branch found here, however, this later expansion does not expose the runaway region to the parent universe. The entire $V<0$ region has already been enclosed before the bifurcating horizon develops. The supercritical evolution therefore gives a geometrically richer
realisation of gravitational cloaking than the subcritical evolution, despite the two branches sharing the same final outcome for the parent universe being an ordinary primordial black hole embedded in the radiation dominated exterior. It is useful to place these two outcomes in the broader context of subcritical
and supercritical vacuum-bubble evolution. The present simulations clearly realise two limiting possibilities: the surviving positive potential barrier is engulfed before it can establish a sufficiently strong locally inflating region, or it becomes sufficiently extended and self-gravitating to generate the
Type-B/child-universe branch. This comparison also suggests a possible intermediate regime in which the barrier becomes locally vacuum dominated and inflates transiently but remains subcritical with respect to the wormhole transition. Whether such configurations disperse, collapse, or are ultimately
re-engulfed by the pre-existing central black hole is not established here and provides a natural target for a future phase-space study.

The dynamically generated bifurcating horizon also provides an
interesting connection with the classification of primordial black
hole geometries. In the standard PBH literature, Type-I and Type-II
classify the initial fluctuation according to whether the areal radius is monotonic or already contains a stationary point, whereas Type-A and Type-B classify the trapping-horizon structure produced during the subsequent collapse \cite{Kopp2011,Uehara_2025,shimada2024primordialblackholeformation}.
These classifications are often closely associated: Type-I initial
fluctuations are normally connected with Type-A PBHs, while the
bifurcating horizons of the Type-B branch have been studied in the
context of sufficiently large Type-II fluctuations. In a dust system the two classifications can in fact be shown to coincide, although this correspondence need not hold in the presence of pressure. Our evolution provides a particularly clear example in which the initial and final classifications do not coincide. The initial hypersurface has $R(t_0,r)=r$ and is therefore
Type-I-like, with no initial throat or non-monotonicity of the areal radius. The collapse of the negative potential region first produces an ordinary Type-A apparent horizon. Only much later does the surviving positive potential barrier deform the exterior geometry sufficiently to generate a minimum of $R(r,t)$ and the bifurcating trapping horizon characteristic of a Type-B PBH. Thus a Type-B geometry need not be encoded in a Type-II initial fluctuation as we show it can be generated dynamically from Type-I initial data by the subsequent matter evolution. A closely analogous dynamical generation of a throat occurs in the
supercritical domain wall evolutions discussed in Refs.~\cite{Deng:2016vzb,Deng:2017uwc} where the initial configuration does not contain a wormhole throat but the subsequent self-gravity of a sufficiently large wall generates a wormhole and baby-universe branch.

An important question is how this picture translates to the Standard Model Higgs. The simulations performed here use a Higgs-inspired metastable potential and vary the overall scalar scale together with the cosmological scale, keeping ratios such as $h_{\max}/H_i$ fixed. For the physical Higgs, the potential and hence $h_{\max}$ and $V_{\max}$, are fixed by particle physics, while the inflationary scale is an independent quantity. A fully realistic Standard Model calculation has not been carried out in the present work. Such a calculation would require fixing the physical Higgs potential while varying $H_i$, the fluctuation amplitude and the spatial lengthscale independently. In particular, the strongly self-gravitating branch is pushed towards increasingly extended superhorizon configurations as the inflationary scale is lowered, making the required numerical domains and evolution times substantially larger. We therefore restrict the present study to establishing the nonlinear mechanisms using a Higgs-like potential and leave a dedicated Standard Model parameter study for future work. This changes the physics in an interesting way. For a fixed Higgs potential, the initial importance of the barrier relative to the post-reheating radiation scales approximately as
\begin{equation}
    \frac{V_{\max}}{\rho_{ \rm fluid}(t_0)}
    \simeq
    \frac{8\pi G V_{\max}}{3H_i^2}.
\end{equation}
Lowering $H_i$ therefore makes a fixed Higgs barrier increasingly important relative to the radiation background. At the same time however, the stochastic fluctuations responsible for pushing the Higgs over the barrier have characteristic amplitude
\begin{equation}
    \delta h\sim\frac{H_i}{2\pi},
\end{equation}
so sufficiently low inflationary scales make the required over-the-barrier excursions increasingly difficult to produce. The realistic Standard Model problem involves a competition between production probability and nonlinear gravitational importance. Lowering $H_i$ suppresses the typical stochastic kick, making over-the-barrier excursions harder to generate, while for a fixed Higgs potential it simultaneously increases the relative importance of the barrier energy compared with the radiation density. A dedicated Standard Model calculation must therefore determine not only how frequently such patches are produced, but also where the resulting profiles lie within the subcritical/supercritical phase structure identified here.

The spatial scale of the fluctuation introduces a second independent effect. The representative evolutions studied in the main text use $H_i r_L=20$, whereas a stochastic Higgs fluctuation generated many e-folds before the end of inflation can correspond to a substantially larger superhorizon region. During radiation domination,
\begin{equation}
    R_L\propto a,
    \qquad
    H^{-1}\propto a^2,
\end{equation}
and hence
\begin{equation}
    H R_L\propto a^{-1}.
\end{equation}
A larger initial fluctuation therefore remains outside the horizon for longer. For an initial size $H_ir_L=q$, horizon entry occurs approximately at
\begin{equation}
    \frac{a_{ \rm HC}}{a_i}\sim q,
    \qquad
    \frac{t_{ \rm HC}}{t_i}\sim q^2 .
\end{equation}
At the same time, increasing the spatial scale increases the physical radius $R_w$ of the barrier and therefore changes its self-gravity. The late-time branch of a physical Higgs fluctuation can consequently depend on its lengthscale as well as its field amplitude and the inflationary energy scale. Our simulations with larger $r_L$ show that the first collapse into a
primordial black hole remains robust as the initial fluctuations are extended. A systematic study of the subsequent branch structure as a function of $r_L$ remains an important next step. In particular, stochastic inflation produces a distribution of amplitudes and lengthscales rather than a single profile, so there is no reason to expect the physical Higgs population to lie exclusively in either the subcritical or supercritical regime. A calculation with the fixed Standard Model potential should vary $H_i$, the fluctuation amplitude and $r_L$ independently and determine which part of this stochastic distribution satisfies
\begin{equation}
    \frac{V}{\rho_{ \rm fluid}}\gtrsim1,
    \qquad
    \frac{R_w}{\ell_{\sigma, \rm eff}}\gtrsim1 .
\end{equation}

The cosmological significance of the two branches is nevertheless the same. The bifurcating horizon distinguishes two very different intermediate geometries, rather than separating a destructive outcome from a safe one. Across the black hole forming configurations studied here, the negative potential region ultimately becomes causally hidden from the parent radiation dominated universe. Once the scalar configuration has either been accreted or disconnected through the supercritical branch, the parent universe contains an ordinary primordial black hole whose subsequent growth is controlled by accretion from the radiation background. Over the time interval followed in our simulations,
the apparent horizon continues to grow, consistent with ongoing accretion from the surrounding radiation fluid. As the cosmological radiation density redshifts, this accretion is expected to become progressively less efficient, so that the black hole mass should approach a finite asymptotic value at sufficiently late times. We do not evolve the simulations sufficiently far to determine this asymptotic mass quantitatively.

Determining what this implies quantitatively for Higgs metastability then requires combining the nonlinear evolution found here with the stochastic probability of producing the corresponding initial configurations. Such a calculation would determine how frequently the two branches occur, the resulting primordial-black hole mass distribution and, ultimately, whether the post-reheating collapse channel modifies existing bounds on the inflationary scale. Extending the simulations beyond spherical symmetry is the other essential step, since the robustness of both the initial cloaking and the later bifurcating horizon geometry must ultimately be established for generic inflationary fluctuations.

\section{Conclusions}

We have used fully nonlinear, spherically symmetric numerical relativity to study the post-inflationary evolution of superhorizon fluctuations of a metastable Higgs-like spectator field in a radiation dominated universe. Across the black hole forming configurations investigated here, we find a robust two stage evolution. The runaway negative potential region first reverses its expansion, enters a kinetic dominated contraction phase and forms a primordial black hole. The subsequent evolution is then controlled by the positive potential barrier that survives outside this first horizon.

Two qualitatively different late-time branches arise. In the subcritical branch, the original apparent horizon grows smoothly outwards and progressively engulfs the remaining scalar configuration. In the supercritical branch, the surviving barrier becomes sufficiently self-gravitating and locally potential dominated to reorganise the exterior spacetime, dynamically generating a bifurcating trapping horizon, a transient wormhole throat and an expanding child-universe branch. Despite these very different intermediate geometries, the outcome for the parent radiation dominated universe is the same: the entire negative potential region becomes causally hidden behind horizons, leaving an ordinary primordial black hole in the exterior. The bifurcating horizon therefore distinguishes two different gravitational evolutions rather than a destructive and a safe outcome.

The present work establishes the nonlinear endpoints and identifies the physical quantities controlling the post-reheating evolution of Higgs-like over-the-barrier patches. The next step is a quantitative application to the physical Standard Model Higgs, combining the nonlinear dynamics identified here with the  stochastic distribution of fluctuation amplitudes and lengthscales. Even when the surviving Higgs-like barrier becomes supercritical enough to generate a transient wormhole and child-universe branch, nonlinear gravity keeps the runaway core cloaked and leaves the parent universe with an ordinary primordial black hole.

\section{Acknowledgments}
EM would like to thank Katy Clough for many useful discussions. EM is supported by an STFC studentship. L.E.P. acknowledges support from JSPS KAKENHI Grant Number JP25KF0278 and UNAM-PAPIIT grant IN114626, ‘Estudios de ANPs y Contribuciones a la colaboración LSST-México.'  DJM acknowledges financial support from the STFC under grant ST/X000931/1.

\appendix\label{sec:appendix}
\section{Numerical accuracy}
As a diagnostic of the numerical accuracy of the simulations, we monitor the constraint relation obtained from the radial derivative of the Misner-Sharp mass. From the Einstein equations,

\begin{equation}
M^\prime = 4\pi R^2\left(R^\prime T_{00} - \dot{R} T_{01}\right),
\end{equation}

which corresponds to a combination of the Hamiltonian and momentum constraints in the variables evolved here. We therefore define the constraint violation as

\begin{equation}\label{eq:Hconstraint}
\mathcal{H}
\equiv
M^\prime - 4\pi R^2\left(R^\prime T_{00} -\dot{R} T_{01}\right).
\end{equation}

For an exact solution of the Einstein equations, $\mathcal{H}=0$. A non-zero value in the numerical solution therefore measures the error introduced by the spatial discretisation, time integration, interpolation and, after black hole formation, the treatment of the excision boundary.

The absolute magnitude of $\mathcal{H}$ is not by itself an ideal measure of the accuracy because the individual terms entering Eq.~\eqref{eq:Hconstraint} can vary by many orders of magnitude during the collapse. In particular, this occurs as strong gradients develop close to the black hole. We therefore also define the normalised residual
\begin{equation}
\frac{|\mathcal{H}|}
{\displaystyle\sum_i |E_i|},
\label{eq:Res}
\end{equation}

where $E_i$ denotes the individual terms appearing in Eq.~\eqref{eq:Hconstraint}. This measures the failure of the cancellation required by the constraint relative to the local magnitude of the terms themselves. Values much smaller than unity therefore indicate good relative constraint satisfaction. As shown below, away from the excision boundary the residual remains many orders of magnitude below unity throughout the exterior domain.

\begin{figure}[h!]
    \centering
    \includegraphics[width=0.8\linewidth]{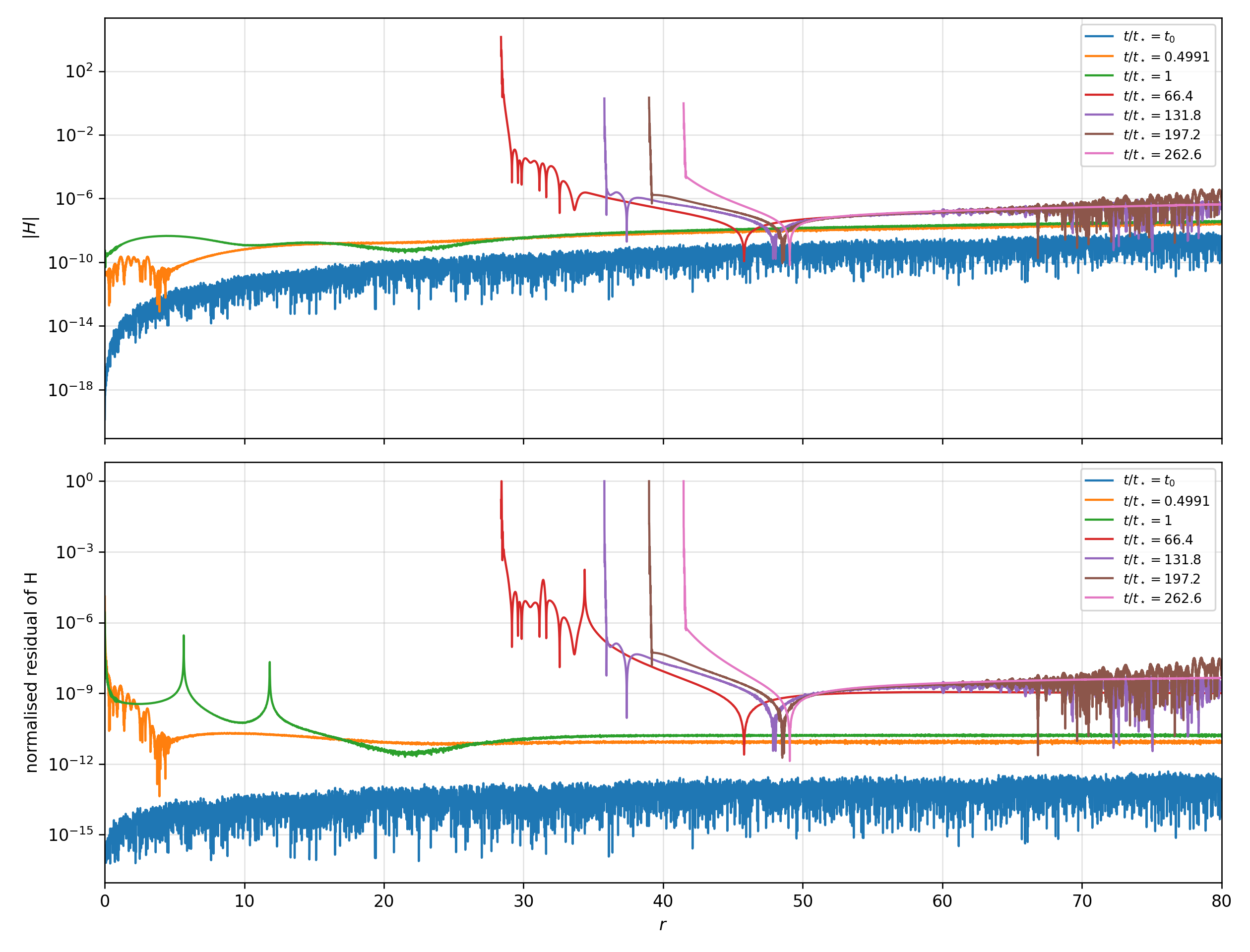}
    \caption{\footnotesize{Constraint diagnostics during the supercritical evolution. Top: absolute residual $|\mathcal H|$ of the Misner--Sharp constraint, Eq.~\eqref{eq:Hconstraint}, on a sequence of time slices. Bottom: the corresponding residual normalised by the sum of the absolute magnitudes of the individual terms entering the constraint.}}
    \label{fig:H_super}
\end{figure}

\begin{figure}[h!]
    \centering
    \includegraphics[width=0.8\linewidth]{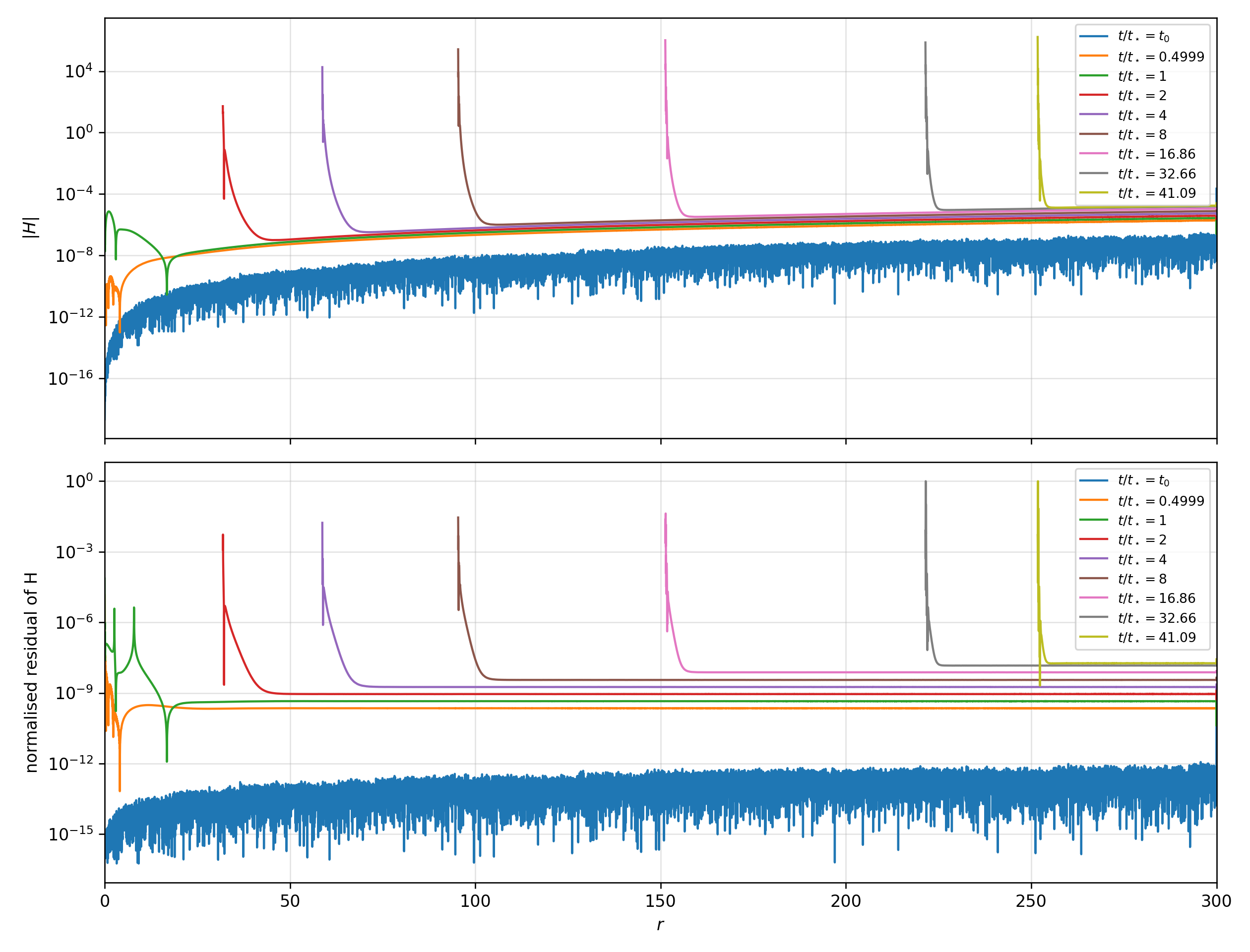}
    \caption{\footnotesize{Constraint diagnostics during the subcritical evolution. Top: absolute residual $|\mathcal H|$ of the Misner--Sharp constraint, Eq.~\eqref{eq:Hconstraint}, on a sequence of time slices. Bottom: the corresponding residual normalised by the sum of the absolute magnitudes of the individual constraint terms.}}
    \label{fig:H_sub}
\end{figure}

After the formation of the first apparent horizon, the region interior to the outermost marginally trapped surface is excised to avoid the high curvature interior terminating the evolution. The excision boundary is moved outward as the horizon grows, requiring interpolation onto the new grid and one-sided finite-difference stencils near the inner boundary. This can produce localised increases in $|\mathcal H|$ and its normalised residual at the first few retained grid points since they may exist behind the horizon during excision. Away from the excision boundary, the normalised residual remains many orders of magnitude below unity, indicating that the constraint remains well satisfied throughout the exterior domain. Figures ~\ref{fig:H_super} and \ref{fig:H_sub} show the constraint violation and the residual for the case of the supercritical and subcritical evolution at time slices across the simulation, respectively.

\subsection{Resolution dependence of the supercritical bifurcation}

Since the bifurcating trapping horizon is a central feature of the
supercritical evolution and develops after the excision procedure has begun, we test its sensitivity to the numerical resolution. We repeat the
representative supercritical simulation at four resolutions,
$N=9000$, $11000$, $13000$, and $15000$, keeping the physical initial conditions and remaining numerical parameters fixed.

\begin{figure}[!ht]
    \centering
    \includegraphics[width=0.85\linewidth]{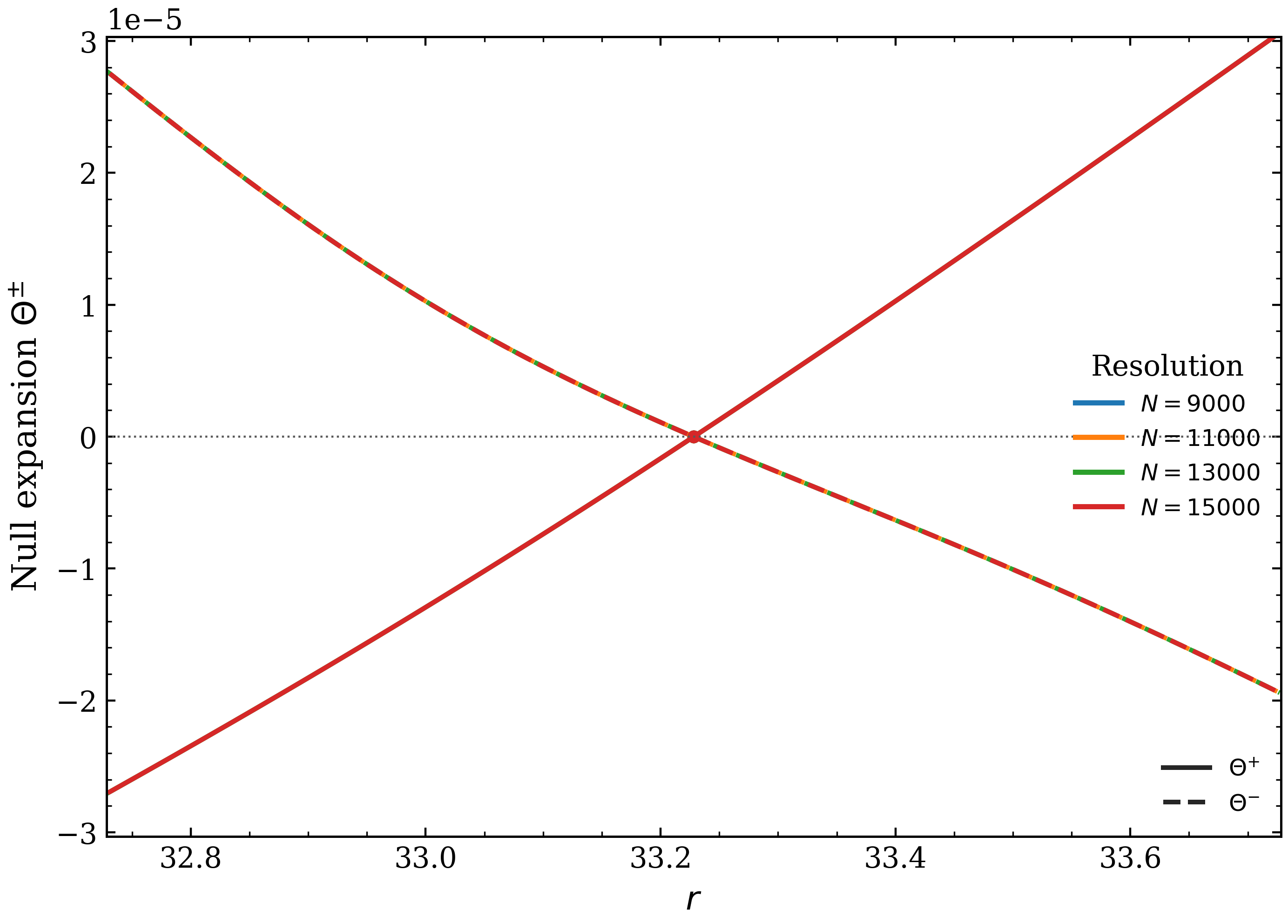}
    \caption{\footnotesize{
    Resolution dependence of the null expansions in the neighbourhood of the bifurcating trapping horizon for the representative supercritical  evolution. Solid and dashed curves show $\Theta^{+}$ and $\Theta^{-}$,  respectively, for resolutions $N=9000$, $11000$, $13000$, and $15000$.
    The curves overlap at the scale shown and the simultaneous zero of $\Theta^{+}$ and $\Theta^{-}$ occurs at the same radial position,
    demonstrating the numerical stability of the secondary bifurcating
    horizon.
    }}
    \label{fig:bif_resolution}
\end{figure}

Figure~\ref{fig:bif_resolution} compares the outgoing and ingoing null expansions in the neighbourhood of the bifurcating horizon. The profiles are indistinguishable at the scale shown for all four resolutions. In each case, the two null expansions vanish simultaneously at the same location, demonstrating that the secondary bifurcating trapping horizon is not a resolution dependent feature of the evolution.

The corresponding bifurcation time and coordinate radius are listed in Table~\ref{tab:bif_resolution}. Both quantities are stable under increasing resolution. The bifurcation time is unchanged to the quoted precision, $t_{ \rm bif}=3543.517$, while the coordinate location changes by only $\sim10^{-6}$ across the resolutions considered.

\begin{table}[t]
    \centering
    \caption{
    Resolution dependence of the bifurcation time $t_{ \rm bif}$ and
    coordinate radius $r_{ \rm bif}$ at which
    $\Theta^{+}=\Theta^{-}=0$ in the representative supercritical evolution.
    }
    \label{tab:bif_resolution}
    \begin{ruledtabular}
    \begin{tabular}{ccc}
        $N$ & $t_{ \rm bif}$ & $r_{ \rm bif}$ \\
        \hline
        9000  & 3543.517 & 33.228188 \\
        11000 & 3543.517 & 33.228189 \\
        13000 & 3543.517 & 33.228189 \\
        15000 & 3543.517 & 33.228189 \\
    \end{tabular}
    \end{ruledtabular}
\end{table}

The agreement between the null expansion profiles and the stability of $(t_{ \rm bif},r_{ \rm bif})$ show that the formation and location of the secondary Type-B trapping-horizon structure are robust against the change in spatial resolution over the range tested.

\section{Dependence on the initial perturbation lengthscale}
\label{app:lengthscale}

The representative simulations in the main text use $r_L/R_H=20$.
To test whether the common first collapse behaviour depends sensitively on this choice, we repeated the evolution for
$r_L/R_H=20,40,80,$ and $160$, keeping the remaining initial profile parameters fixed.

\begin{figure*}[!ht]
    \centering
    \includegraphics[width=0.8\textwidth]{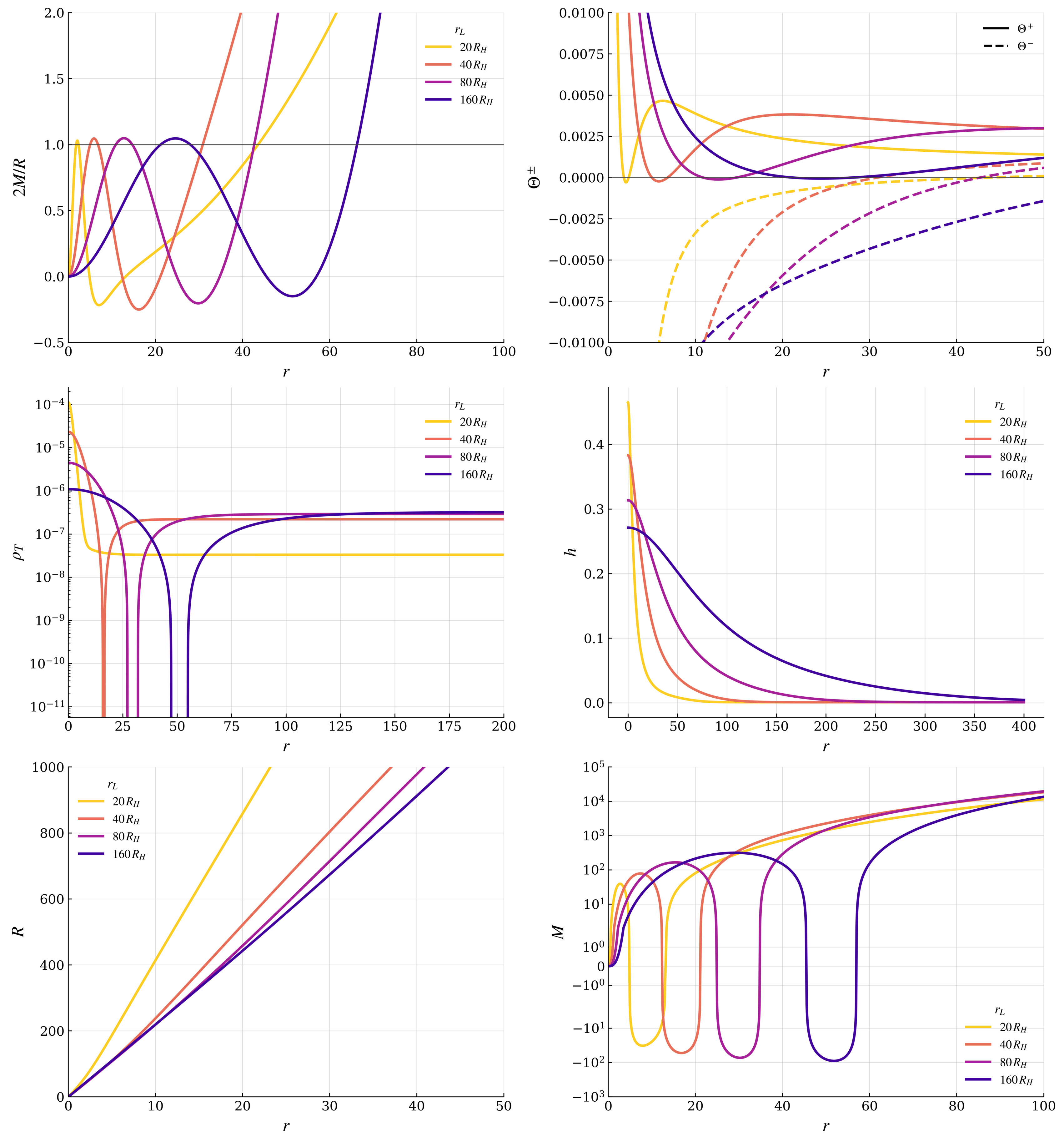}
    \caption{\footnotesize{
    Comparison of the geometric and matter diagnostics at the first
    apparent horizon formation time $t_\star$ for Gaussian initial
    perturbations with $r_L/R_H=20,40,80,$ and $160$.
    \textbf{Top left}: compactness $2M/R$, with the horizontal line marking
    the apparent horizon condition $2M/R=1$.
    \textbf{Top right}: outgoing $\Theta^+$ (solid) and ingoing
    $\Theta^-$ (dashed) null expansions.
    \textbf{Middle left}: total energy density $\rho_T$.
    \textbf{Middle right}: Higgs profile $h$.
    \textbf{Bottom left}: areal radius $R$.
    \textbf{Bottom right}: Misner--Sharp mass $M$.
    In all four cases the first apparent horizon satisfies
    $\Theta^+=0$ with $\Theta^-<0$; increasing $r_L$ shifts the collapse to larger spatial scales.} }
    \label{fig:comparison_AH}
\end{figure*}

Figure~\ref{fig:comparison_AH} compares the four simulations on the time slice at which the first apparent horizon forms. In every case the compactness reaches $2M/R=1$ and the outgoing null expansion vanishes while the ingoing expansion remains negative, $\Theta^+=0$ and $\Theta^-<0$. Thus the first nonlinear outcome remains the formation of a black hole apparent horizon across the full range of lengthscales considered. Increasing $r_L$ primarily changes the spatial scale of the collapse. The apparent horizon forms at larger radius for more extended initial fluctuations, while the Higgs-field and Misner-Sharp mass profiles extend
over a correspondingly larger radial region. The qualitative robustness of the first collapse therefore does not rely on the particular choice $r_L/R_H=20$ used for the representative figures in the main text. 

\bibliography{bib}

\end{document}